%% file: main.tex
\newif\ifchange

\documentclass[sigconf, usenames]{acmart}
\usepackage{xcolor}
\newcommand{\change}[1]{\textcolor{black}{#1}}
\usepackage{longtable}
\usepackage{dblfloatfix} 
\usepackage{booktabs} 
\usepackage{arydshln}
\usepackage{subfig}
\usepackage{url}
\usepackage{hyperref}
\copyrightyear{2025}
\acmYear{2025}
\setcopyright{acmlicensed}\acmConference[UIST '25]{The 38th Annual ACM Symposium on User Interface Software and Technology}{September 28-October 1, 2025}{Busan, Republic of Korea}
\acmBooktitle{The 38th Annual ACM Symposium on User Interface Software and Technology (UIST '25), September 28-October 1, 2025, Busan, Republic of Korea}
\acmDOI{10.1145/3746059.3747664}
\acmISBN{979-8-4007-2037-6/2025/09}

\AtBeginDocument{%
  \providecommand\BibTeX{{%
    \normalfont B\kern-0.5em{\scshape i\kern-0.25em b}\kern-0.8em\TeX}}}

\newcommand{\system}{MapStory}

\begin{document}
\title{\system{}: Prototyping Editable Map Animations \\ with LLM Agents}





\author{Aditya Gunturu}

\affiliation{%
  \institution{University of Calgary}
  \country{Canada}
  }
\affiliation{%
  \institution{CU Boulder}
  \country{United States}
  }
\email{aditya.gunturu@ucalgary.ca}

\author{Ben Pearman}
\affiliation{%
  \institution{University of Calgary}
  \country{Canada}
  }
\email{ben.pearman@ucalgary.ca}

\author{Keiichi Ihara}
\affiliation{%
  \institution{CU Boulder}
  \country{United States}
  }
\email{keiichi.ihara@colorado.edu}

\author{Morteza Faraji}
\affiliation{%
  \institution{University of Calgary}
  \country{Canada}
  }
\email{morteza.faraji@ucalgary.ca}

\author{Bryan Wang}
\affiliation{%
  \institution{Adobe Research}
    \country{United States}
  }
\email{bryanw@adobe.com}

\author{Rubaiat Habib Kazi}
\affiliation{%
  \institution{Adobe Research}
    \country{United States}
  }
\email{rhabib@adobe.com}

\author{Ryo Suzuki}
\orcid{0000-0003-3294-9555}
\affiliation{%
  \institution{CU Boulder}
  \country{United States}
  }
\email{ryo.suzuki@colorado.edu}

\renewcommand{\shortauthors}{Gunturu, et al.}
\input{0-abstract}

\begin{teaserfigure}
\includegraphics[width=1\textwidth]{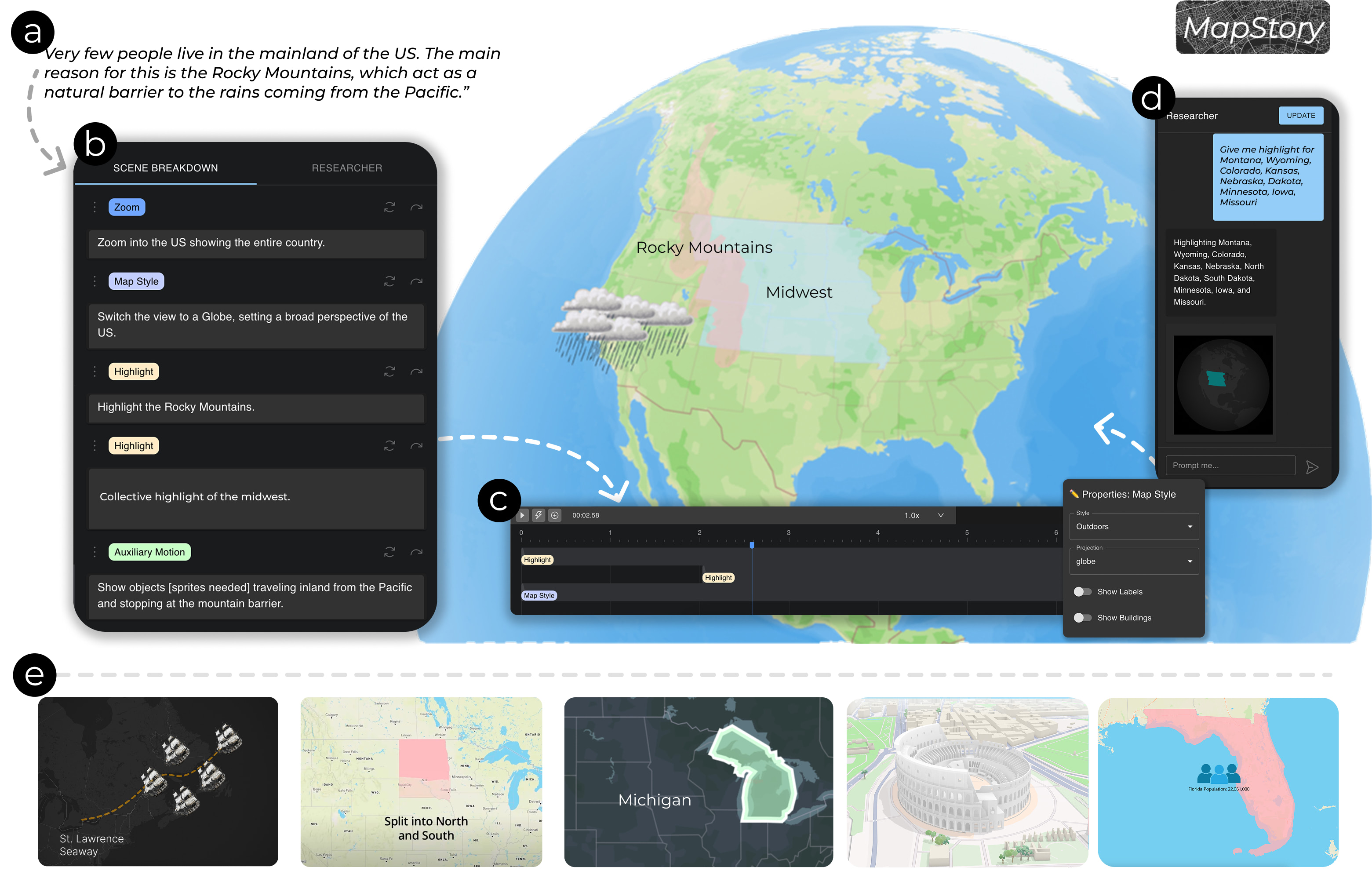}
\centering
\caption{\system{} is an LLM-powered tool for prototyping map-centric animations from natural language scripts. Users write and edit a script (a), which generates an editable scene breakdown (b). The animation is shown on a timeline with a properties panel (c), and a researcher module retrieves and updates geospatial data (d). We showcase various map animation examples (e).}
\label{fig:teaser}
\end{teaserfigure}

\maketitle

\input{1-introduction}

\input{2-related-work}
\input{3-formative-study}
\input{4-taxonomy}

\input{5-system}

\input{6-implementation}

\input{7-user-study}

\input{8-future-work}

\input{9-conclusion}
\input{acknowledgements}

\balance
\bibliographystyle{ACM-Reference-Format}
\bibliography{references}

\input{videosource}
\end{document}

%% file: 0-abstract.tex
\begin{abstract}
We introduce MapStory, an LLM‑powered animation prototyping tool that generates editable map animation sequences directly from natural language text by leveraging a dual-agent LLM architecture. Given a user-written script, MapStory automatically produces a scene breakdown, which decomposes the text into key \change{map animation primitives} such as camera movements, visual highlights, and animated elements. Our system includes a researcher agent that accurately queries geospatial information by leveraging an LLM with web search, enabling automatic extraction of relevant regions, paths, and coordinates while allowing users to edit and query for changes or additional information to refine the results. Additionally, users can fine-tune parameters of these primitive blocks through an interactive timeline editor. We detail the system’s design and architecture, informed by formative interviews with professional animators and by an analysis of 200 existing map animation videos. Our evaluation, which includes expert interviews (N=5), and a usability study (N=12), demonstrates that MapStory enables users to create map animations with ease, facilitates faster iteration, encourages creative exploration, and lowers barriers to creating map-centric stories.


\end{abstract}

\ccsdesc[500]{Human-centered computing~User interface design}  

\keywords{text-to-animation, map-based storytelling, LLM-based authoring tools, AI-assisted animation, human-AI collaboration}

%% file: 1-introduction.tex
\begin{figure*}
\includegraphics[width=1\textwidth]{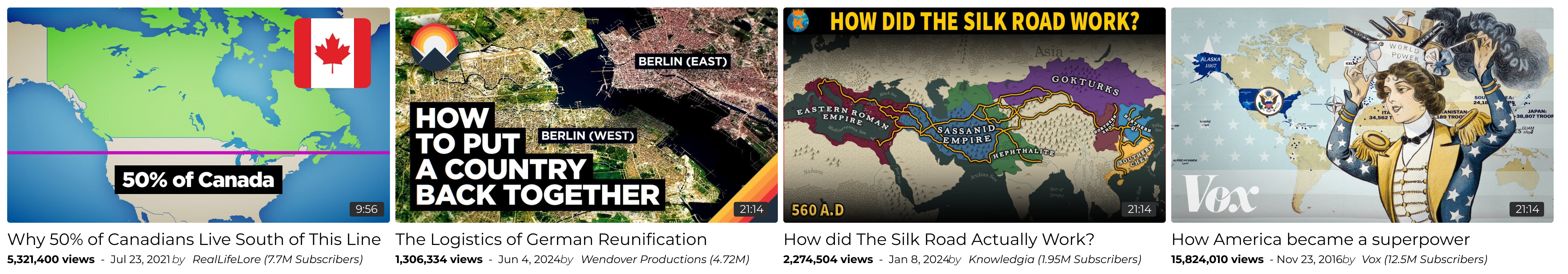} 
  \caption{Popular online animated map videos explain knowledge to a general audience in a highly
engaging manner, influencing millions of viewers (See list of image sources for credits and links).}
  \label{fig:examplevideos}
\end{figure*}

\section{Introduction}

Maps have long been a powerful medium for storytelling~\cite{wood1992power}. Since ancient and medieval times, people have used maps to tell stories~\cite{tufte1991envisioning, tufte1983visual}, such as the Catalan Atlas, which depicted medieval trade routes~\cite{Catalan_Atlas}, or Charles Minard’s visualization of Napoleon’s campaign, considered one of the earliest examples of map-based data storytelling~\cite{Charles_Minard}. Today, people bring this form of narrative to life through animated storytelling, which we call \textit{\textbf{map animation}}. This map animation is becoming increasingly popular, as it can effectively communicate historical trajectories~\cite{unknown_history_2021}, migration patterns~\cite{real_life_core_canadians_south_of_line}, or geopolitical dynamics~\cite{knowledgia_silk_2022}, often attracting millions of viewers as engaging educational content.

However, creating map animations remains significantly challenging and time-consuming. Producing even a short animation clip requires substantial efforts and expertise with animation tools. In addition, creators must conduct extensive research to ensure factual accuracy throughout the process. As a result, the creation of map animations remains largely inaccessible to non-experts, and even professional animators often spend days or weeks producing a single video, as we observed in our formative study. \change{While not all users may seek to create animated maps, existing apps like TravelAnimator\cite{travelanimator} and many popular YouTube tutorials available online \cite{tutorial1, tutorial2, tutorial3}  indicate clear novice interest in editable map animation creation.}

In this paper, we ask: \textit{what if anyone could immediately create these map-based animations simply by writing a script}? And going beyond that to make them editable at every stage. For example, imagine typing a sentence like \textit{``Very few people live in the mainland of the US. The main reason for this is the Rocky Mountains, which act as a natural barrier to the rains coming from the Pacific''} and instantly generating an animated sequence that visualizes this explanation and is editable at the script level as well as granular animation level (Figure~\ref{fig:teaser}). To explore this vision, we present \textbf{\system{}} \footnote{Project Page: \href{https://adigunturu.github.io/MapStory-UIST25/}https://adigunturu.github.io/MapStory-UIST25/}, a system that generates editable map animations from natural language text following a human-in-the loop approach throughout the process. Unlike general-purpose text-to-video models like Sora~\cite{brooks2024video} or Veo~\cite{veo2}, \system{} is designed to preserve creative control and enable real-time fine-grained editing through \textit{a scene breakdown}, a \textit{conversational researcher}, and \textit{primitive map animation blocks} such as camera motion, location highlights, and animated elements presented in an editable timeline interface. \change{Our goal is not to replace traditional professional tools like After Effects and their low‑level functions. Instead, we provide a \textit{prototyping environment} that lets animators quickly draft map animations from text and refine them through rapid iteration with parametric control, enabling faster exploration and ideation.}


MapStory is powered by an agentic architecture consisting of two LLM-based agents: 1) \textbf{\textit{a scene breakdown agent}}, which parses the script into map animation primitives, and 2) \textbf{\textit{a researcher agent}}, which grounds the script in factual geospatial data by identifying real-world regions, paths, and coordinates. For instance, in the previous Rocky Mountains example (Figure~\ref{fig:teaser}a), the scene breakdown agent identifies zooming actions for the US, highlights for the Rockies and Midwest regions, and animated elements like moving clouds (Figure~\ref{fig:teaser}b). Furthermore, the researcher agent automatically fetches polygons to highlight these regions without requiring manual specification; the user may then provide additional modification queries (Figure \ref{fig:teaser}d). These animations can be easily modified or fine-tuned through an interactive timeline interface (Figure~\ref{fig:teaser}c). We leverage the geocoding ability of LLMs \cite{manvi2023geollm, roberts2023gpt4geo} to generate real‑world Cartesian coordinates and implement our architecture based on OpenAI’s \textit{o1} model for the scene breakdown agent and Perplexity’s \textit{sonar‑pro} model for the researcher agent with web‑search capability.

Our design was informed by two formative studies. First, we identified \textit{ animation primitives} commonly used in map animations (Figure~\ref{fig:teaser}e and \ref{fig:taxonomy}) through an analysis of 200 publicly available map animation videos on YouTube. These primitive animations fall into three categories: 1) \textbf{\textit{camera movements}}, such as \textit{zoom}, \textit{translate}, and \textit{orbit}; 2) \textbf{\textit{visual highlights}}, such as \textit{area}, \textit{path}, and \textit{point highlights}; and 3) \textbf{\textit{animated elements}}, such as \textit{routes}, \textit{spatial transitions}, and \textit{auxiliary motion}. Second, we interviewed professional animators to understand their workflows and challenges, informing our three key design goals: 1) script-driven authoring, 2) research integration, and 3) modular map animation blocks for creative controllability. These findings guided the design and development of the \system{} interface and system architecture.

To evaluate our system, we conducted three studies: 1) \textit{a usability study} (N=12) to measure the tool’s expressiveness, creative exploration, and accessibility for novice users; 2) \textit{an expert study} with professional map animators (N=5) to gather feedback on the tool's workflows, and potential for real-world applicability; and 3) \textit{a technical evaluation} on 20 randomly sampled prompts to assess the generalizability and factual correctness of automatically generated animations. Our findings suggest that \system{} enables faster iteration, encourages creative exploration, and lowers barriers to creating map-centric animations for storytelling. Participants appreciated the effective balance between automation and control, and experts recognized its potential as a creative support tool with a focus on rapid prototyping. On the other hand, we also identified several limitations, including challenges with AI hallucination affecting factual accuracy the need to verify produced visualizations.

Finally, our contributions are as follows:
\begin{enumerate}
\item \system{}, a text-driven animation prototyping tool powered by an LLM-based agent architecture, composed of scene breakdown and researcher agents, that enables automated yet human-in-the-loop creation of map animations. 
\item Insights from two formative studies: analyzing 200 map animation videos and interviewing expert animators, that guided our identification of essential \change{primitive map animation blocks} and script-driven design principles.
\item Results from three evaluations: technical, usability, and expert feedback, that assess the effectiveness of our tool and highlight opportunities and challenges in AI-assisted map animation authoring. 
\end{enumerate}

%% file: 2-related-work.tex
\section{Related Work}
\subsection{Map-Based Visualization and Storytelling}
Map-based storytelling systems have leveraged interactive visualizations, especially geographic maps, to contextualize narratives. For example, NewsViews~\cite{gao2014newsviews} and GeoExplainer~\cite{lei2023geoexplainer} automatically generate annotated thematic maps by mining text for locations and linking them to relevant data. Lundblad et al.~\cite{lundblad2013geovisual} present a web-based geovisual analytics toolkit that integrates dynamic visual analysis with interactive storytelling. These systems lower the barrier to creating map-based storytelling for understanding statistical data and conducting spatial analyses. 
Narrative Maps~\cite{keith2021narrative} introduced a “route map” metaphor, depicting events in a story as landmarks connected by routes on a conceptual map. 
Prior work has extensively explored map-based visualizations for geospatial data. For example, data visualizations rendered on maps or 3D globes to convey global trends~\cite{satriadi2021quantitative, satriadi2022tangible}. Other work integrates narratives with physical space: Location-Aware Adaptation~\cite{li2023location}, Story-Driven~\cite{belz2024story}, Believable Environments~\cite{gustafsson2006believable}, and How Space is Told~\cite{shin2023space} present an approach to generate location-based stories, automatically assigning story events to contextually suitable locations.
Additionally, GeoCamera~\cite{li2023geocamera} introduced an authoring tool that supports users in designing camera movements for storytelling with geographic visualizations. 
MapStory builds on these threads and extends prior map visualization and narrative authoring systems by leveraging map animation generation through script, allowing authors to easily produce and revise rich map-based stories.

\subsection{Animation Authoring Tools}
Creating animations traditionally requires significant time, effort, and technical skill, so a number of HCI systems have explored ways to lower this barrier. Early approaches introduced sketch-based animation tools like K-Sketch~\cite{davis2008k}, Draco~\cite{kazi2014draco}, and Kitty~\cite{kazi2014kitty}. For example, Draco~\cite{kazi2014draco} let illustrators bring static drawings to life by sketching motion paths and applying kinetic textures to create rich path animations and particle effects. These sketch-based interfaces greatly lowered the barrier through more natural interactions. Augmented Physics \cite{gunturu2024augmented} introduced a CV-based pipeline to convert static physics diagrams into animated physics simulations, providing a AI-assisted, selection oriented tool for creating animated physics visuals. Another line of work provides higher-level building blocks and templates to simplify animation authoring. Motion Amplifiers~\cite{kazi2016motion} introduced a set of reusable animation primitives based on the principles of traditional animation. Similarly, Ma et al.~\cite{ma2022layered} extended this idea to 3D animations with a layered authoring interface. This layered design balanced ease-of-use with expressiveness, showing how modular animation blocks and multi-level interfaces can support both novices and professionals. Other systems have explored performance-based and data-driven templates: for example, Moscovich et al.~\cite{moscovich2001animation} enabled recording motions via hand gestures. 
More recently, researchers have explored flowchart-based authoring~\cite{zhang2020flowmatic, chen2021entanglevr}. For instance, FlowMatic~\cite{zhang2020flowmatic} introduced an authoring tool that uses flowcharts to create reactive behaviors of virtual objects.

\subsection{LLM-Powered Authoring Tools}
The emergence of generative AI and large language models (LLMs) has further inspired AI-assisted authoring tools. LLMs have shown promise in supporting creative work like story generation~\cite{chung2022talebrush}. In particular, recent works have explored text-to-video generation, demonstrated through numerous commercially-available tools, including Sora \cite{openai_sora}, Veo2~\cite{Veo2:online}, Runway~\cite{Runway:online}, and Pika~\cite{Pika:online}. These approaches basically generate videos from prompts, but they usually lack iteration and fine-tuning. 

In contrast, the HCI community has explored more interactive and user-in-the-loop approaches. 
Several systems introduce visual authoring techniques, such as graph-based manipulation~\cite{jiang2023graphologue, yan2023xcreation, arawjo2023chainforge}, drag-and-drop interfaces~\cite{masson2024directgpt, brade2023promptify}, and multi-modal prompt refinement ~\cite{wang2024promptcharm}, to allow users to edit and steer the generation process.
Several notable systems have applied these approaches to animation generation.
One example is Katika~\cite{jahanlou2022katika}, an end-to-end system for creating explainer-style motion graphics videos from a natural language script. Alternatively, Spellburst~\cite{angert2023spellburst} uses an LLM to produce p5.js code snippets to generate creative web animations. Similarly, Keyframer~\cite{tseng2024keyframer} explored using GPT-3 to synthesize CSS keyframe animations from design descriptions. LogoMotion~\cite{liu2024logomotion} is a recent system that helps novices animate static logo graphics through visually-grounded code synthesis. Inspired by these prior works, our system also leverages LLMs not to directly output the video content, but instead generate animation components by leveraging LLM-agent architecture for map-based animation generation.

\subsection{Natural Language for Creative Tasks}
Natural language has been extensively explored as a medium for authoring, editing, and organizing content across various creative domains. 
Prior work has shown its effectiveness in video editing and storytelling with text input. For instance, CrossPower~\cite{xia2020crosspower} leverages script-like natural language to organize visuals in videos, while DataParticles~\cite{cao2023dataparticles} supports language-oriented authoring of unit visualizations.

Natural language is not limited to text input; recent systems explore speech as a modality for real-time creative control. 
TakeToons \cite{subramonyam2018taketoons} leverages structured scripts and the actor's facial poses to translate talking animations to a virtual character in real-time. DrawTalking~\cite{rosenberg2024drawtalking}, on the other hand, enables users to add simple motion to sketched objects via speech, and RealityTalk~\cite{liao2022realitytalk} displays relevant graphics based on the user's speech for creating augmented presentations. 

Recently, many works have focused on editing and organizing video content via natural language using LLMs. Works such as LAVE~\cite{wang2024lave} and ExpressEdit~\cite{tilekbay2024expressedit} have explored story level editing video through free-form text. 
Other systems, such as ChunkyEdit~\cite{leake2024chunkyedit}, QuickCut~\cite{truong2016quickcut}, and B-Script~\cite{huber2019b} have explored natural language prompting as a way to make edits, such as trimming or organizing video clips using video and organizing transcripts. 

Natural language has also been used for map-based content creation. For instance, Embark~\cite{sonnentag2023embark} has examined parsing natural language text outlines into plans with routes rendered on maps, effectively turning notes with a structured schema into dynamic documents. Meanwhile, tools like Eviza~\cite{setlur2016eviza} support natural language analytical questions and render the resulting visualization situated on a map. Finally, CrossTalk~\cite{xia2023crosstalk} leverages conversations in online meetings to organize, recommend, and navigate places on shared maps. Script-driven editing tools for video have proven to be promising in allowing creators to have high-level control over their editing, organization, and overall storytelling \cite{leake2024chunkyedit, truong2016quickcut, leake2017computational}. 
Inspired by these ideas, our system supports creating parameterized animations from the script, which are organizable, controllable, and editable at a high level for conveying map-centric stories.

Recent work in HCI has seen the emergence of LLM-driven editing tools such as ~LAVE \cite{wang2024lave} and MoGraphGPT ~\cite{ye2025mographgpt}, which respectively leverage linguistic augmentation for video editing and modular LLMs for interactive scene creation. In contrast to these approaches, our architecture is novel in its explicit focus on map-centric storytelling: it employs a dedicated \textit{planner-researcher} agent that translates free-form user instructions into GeoJSON operations, including querying, addition, reduction, and generation, while being promptable at every step of the architecture. The user can request changes to the scene breakdown as well as the researcher agent. This modular design not only decouples complex geographic reasoning from traditional video or scene editing pipelines but also provides enhanced transparency and user control by allowing real-time refinement of map animations. 

%% file: 3-formative-study.tex
\section{Formative Study}

\change{To better understand current practices and challenges in creating map animations, we conducted a formative study. Our method is informed by prior work  such as k-Sketch \cite{davis2008k} and Draco \cite{kazi2014draco}, which demonstrated that novices benefit significantly from learning and adopting expert-derived strategies. Moreover, cognitive apprenticeship theory \cite{dennen2008cognitive} shows that novices perform best by watching and trying expert strategies. The goal of this study was to gather insights that would guide the design of \system{}, grounded in the real-world experiences, needs, and challenges faced by professionals map animators. }

\subsection{Method}
We recruited three professional map video animators (3 male, ages 21-28). Each participant had at least three years of experience (3-6 years) producing map animation videos, which they regularly upload to YouTube. One participant does client work. We contacted them via emails, Youtube, and Instagram DMs. We conducted semi-structured interviews of approximately 1 hour over Zoom, and compensated each with 15 USD. 

MapStory was intentionally designed using insights from expert animators, following a cognitive apprenticeship model \cite{dennen2008cognitive} to scaffold novice interactions effectively. This design choice aligns with successful prior systems such as K-Sketch \cite{davis2008k} and Draco \cite{kazi2014draco}, which demonstrated that novices benefit significantly from learning and adopting expert-derived strategies.

During the interviews, we explored the experts’ typical workflows, the tools they use, and the challenges or needs they encounter when producing map animations. Although we acknowledge the relatively small number of experts in our study, we intentionally focused on specialized map animators rather than general video creators, thus recruitment was challenging. To complement our interviews and mitigate this limitation, we also analyzed a well-known tutorial on creating map animations, titled \textit{``How I Make My Maps''}~\footnote{\url{https://www.youtube.com/watch?v=GsojLuJpe_0}}. We also analyzed popular creator's Patreon tutorials~\footnote{\url{https://www.patreon.com/posts/behind-scenes-51811597}}. Our findings, drawn from both the expert interviews and the tutorial analysis, are described in the following sections.

\subsection{Insights and Findings}

\subsubsection*{\textbf{Current Tools, Practices, and Workflows}} 
In our study, all experts use Adobe After Effects for video editing. For map-specific graphics, they use an After Effects plugin called GeoLayers~\cite{GeoLayers}, which offers features such as importing real-world map data, customizing map styles, and integrating location-based metadata. In case of area highlights, they use GeoJSON \footnote{\url{https://geojson.org/}} polygons to render the regions on the maps. Despite these capabilities, the experts noted that complex animations still demand considerable manual scripting and keyframing. As a result, they reported that creating a single map animation typically takes from a few days to weeks, with some projects extending beyond. Moreover, the experts emphasized that After Effects usage is only one component of their overall process. Based on their comments and the tutorial, a typical map animation project generally involves four stages: 1) \textbf{\textit{script phase}}, where the idea and narrative are determined; 2) \textbf{\textit{research phase}}, for gathering factual information and extracting data; 3) \textbf{\textit{animation prototyping phase}}, where the script is translated into a rough animation; and 4) \textbf{\textit{iteration phase}}, during which the animation is refined for publication.

\subsubsection*{\textbf{It All Starts with a Script}} 
\begin{figure}[h!]
\includegraphics[width=1\linewidth]{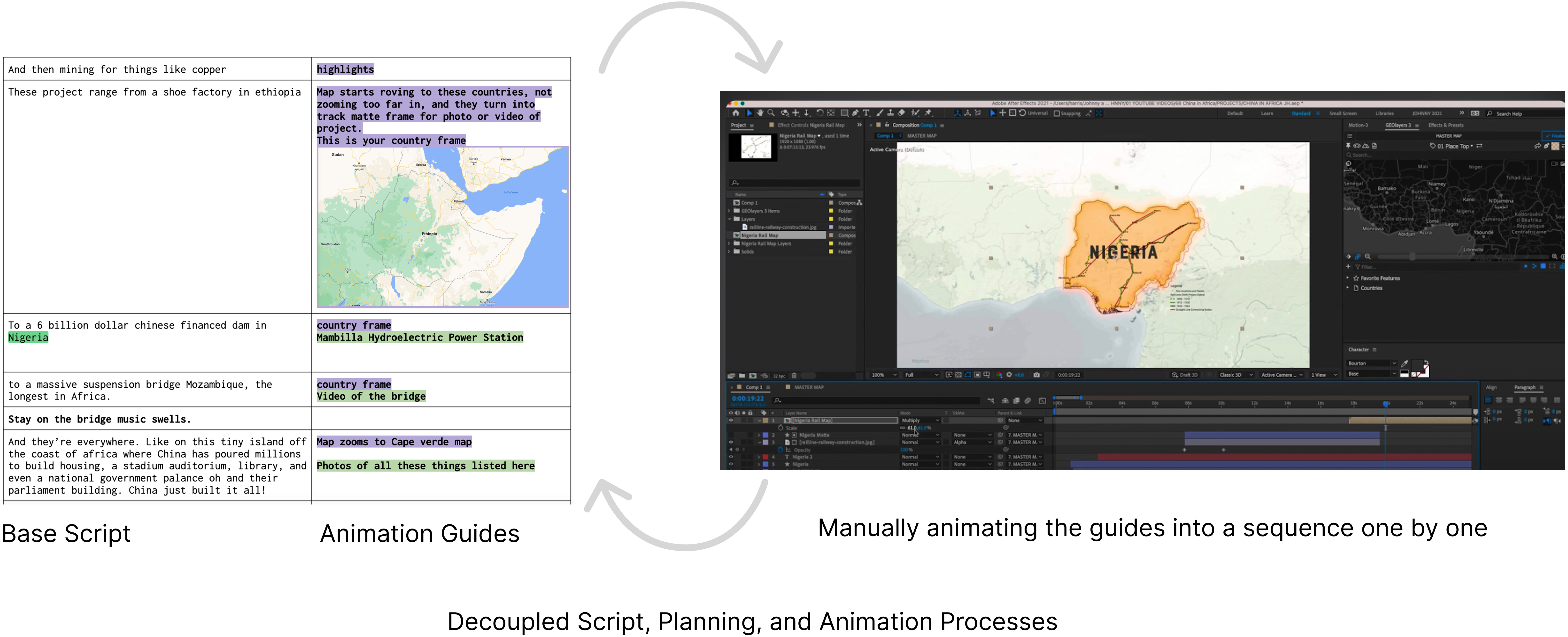} 
\caption{Workflow of a map animator. The animator first breaks down their script into manageable chunks and describes the animation they plan to make for this specific item. The animation guides also include research information like population counts, medals won by a specific country or region boundary maps. ©Patreon - Johnny Harris}
\label{fig:formative_figure}
\end{figure}

Participants emphasized that the animation process always begins with a script that outlines the core narrative, pacing, and visual milestones (Figure~\ref{fig:formative_figure}). Before working in After Effects, they typically create a written plan specifying what content must appear on the map and when, in sync with the video’s narration. As P1 explained, \textit{``I generally start with an idea [...] then I’ll start writing a script, and it usually takes longer than you’d think''}. The script also captures instructions for other visual elements, such as layering different geographic datasets, adding text callouts, or highlighting resource-rich regions. P1 added, \textit{``Sometimes, I’ll put directions like ‘zoom in here’ or ‘highlight this region' in my script''}. Once the script is complete, they conduct a \textit{``script breakdown''} by tagging specific parts of the text with directional notes—such as zooming in or toggling boundaries—to convert the written plan into explicit animation steps. As P2 described, \textit{``When I started, I did some markup on my script like `move the camera here' or `highlight that region next' so I know exactly how to animate it in After Effects''}. Some participants also described using large language models to outline their animation sequences well before launching any software. As P2 explained, \textit{``I give ChatGPT the script and ask it to generate a plan [...] It's a huge time saver for planning my camera moves before I open After Effects.''}

\subsubsection*{\textbf{Decoupled Research and Animation Process}} After writing the script, creators often go through a dedicated research phase to identify accurate routes, verify historical or geographic facts, and gather supporting data. However, this research occurs outside the animation tools, so animators must manually translate their findings into keyframes, such as plotting accurate paths or highlighting specific regions. As P2 noted, \textit{``I do my fact-checking in Google or ChatGPT, but then I have to transfer everything by hand into After Effects. If some information like routes, highlights, or stats changes last minute, I need to rework the whole scene.''} This lack of real-time integration between research, planning, and animation leads to frequent misalignments, increased chances for errors, and slower iteration cycles. Ultimately, these fragmented workflows hinder the creative process. P1 echoed similar challenges, noting how \textit{“I do a lot of the research with Google, but then I basically integrate it all (borders, stats, routes, facts etc.) out by hand. If I discover something’s slightly off, I have to go back and fix every keyframe.”} He also pointed out that \textit{“I’ll sometimes use ChatGPT to refine my script, but there’s no direct way to update the animation when I find new facts—I still have to do everything manually.”} This repeated back-and-forth between gathering accurate data and implementing it in the animation often results in “days” of extra work whenever details change.

\subsubsection*{\textbf{Need for Iteration and Creative Controllability}}

When prototyping their animations, experts emphasized the importance of an iterative process. As P1 explained, \textit{``Being able to quickly tweak an animation and immediately see how it aligns with the narrative is crucial''}. Therefore, rather than relying on a single, monolithic timeline, creators prefer to assemble individual building blocks or primitives, where each block is responsible for a particular sequence or visual effect. For instance, one block might highlight a geographic boundary, while another animates a specific route, which are related to the tags in the script. Since these modules are self-contained, changes in one block do not disrupt the rest of the animation. P2 added, \textit{``We can’t just fix it on the fly in the same place, so it takes forever if the story shifts suddenly.''}. Similarly, P3 said \textit{``When clients request changes last moment, it can be very frustrating. I try to clarify requirements upfront, but clients sometimes change their requests in the middle of the process,''}, underscoring how sudden updates force rework, which takes a lot of time. Overall, professional creators prefer granular-level adjustments and creative controllability, rather than generating the entire video in a single pass.

\begin{figure*}
\includegraphics[width=1\textwidth]{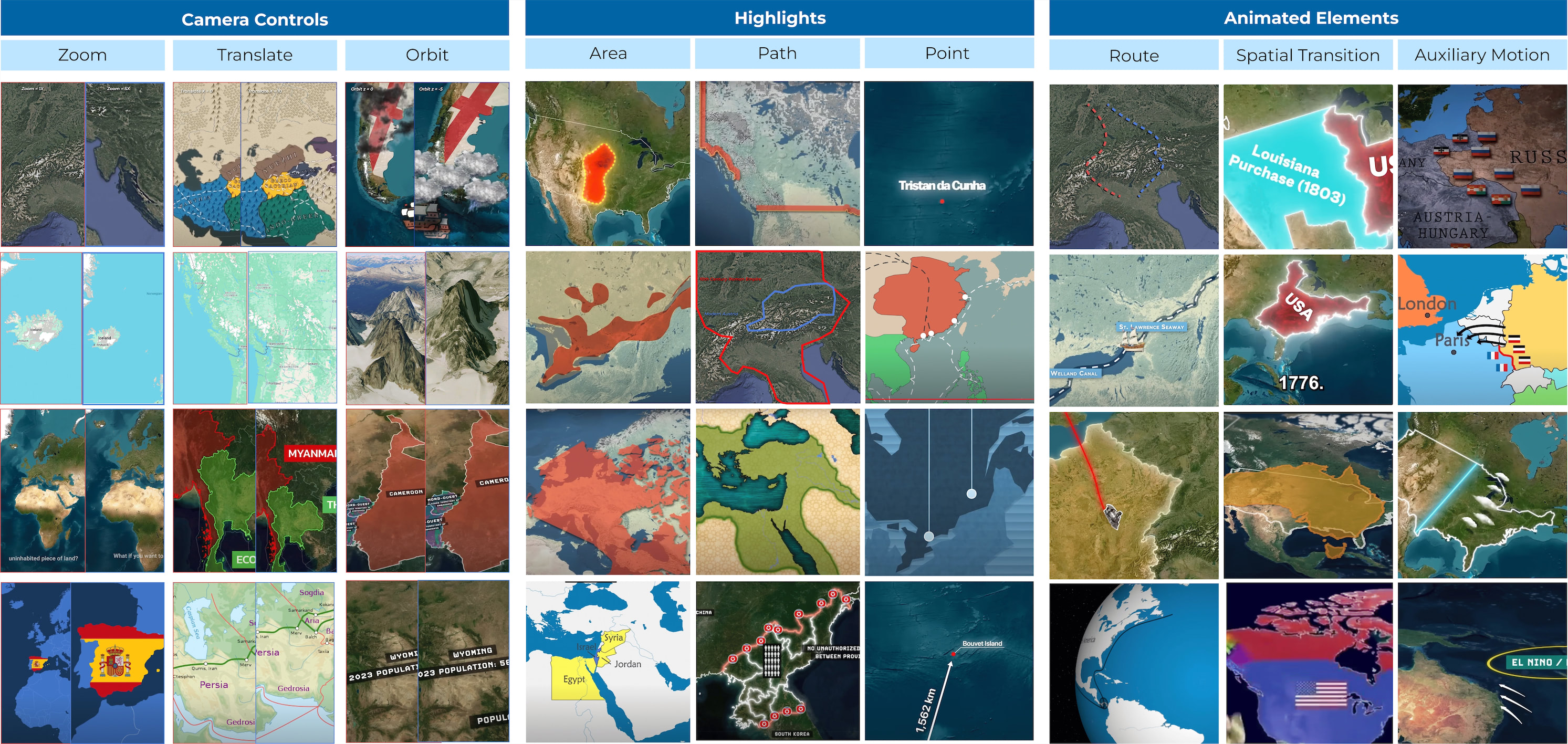} 
  \caption{Common map augmentation and animation techniques based on content analysis of 200 map-centric videos (See list of image sources for credits and links).}
  \label{fig:taxonomy}
\end{figure*}

\subsection{Design Implications}
Based on these findings, we made the following design decisions.

\subsubsection*{\textbf{D1: Script-Driven Authoring}}
Our system should directly translate scripts to support a script-driven workflow. Moreover, changes to textual instructions should seamlessly reflect in the animated map scenes, enabling rapid iteration.

\subsubsection*{\textbf{D2: Research Integration to Extract Map Data}}
Our system should integrate research into the animation tool, allowing users to gather, verify, and organize geospatial data and story elements in a single environment. These elements should be directly embedded into the scene to minimize manual plotting. 

\subsubsection*{\textbf{D3: Generating Editable Animation Building Blocks}}
Our system should generate animations not as one-off videos but as editable animation modules. These modules should allow creators to tweak individual parameters and elements without disrupting other segments, thereby supporting fine-tuned visual adjustments and rapid iteration.

%% file: 4-taxonomy.tex
\section{Common Map Animation Techniques}

\begin{figure*}
\includegraphics[width=1\textwidth]{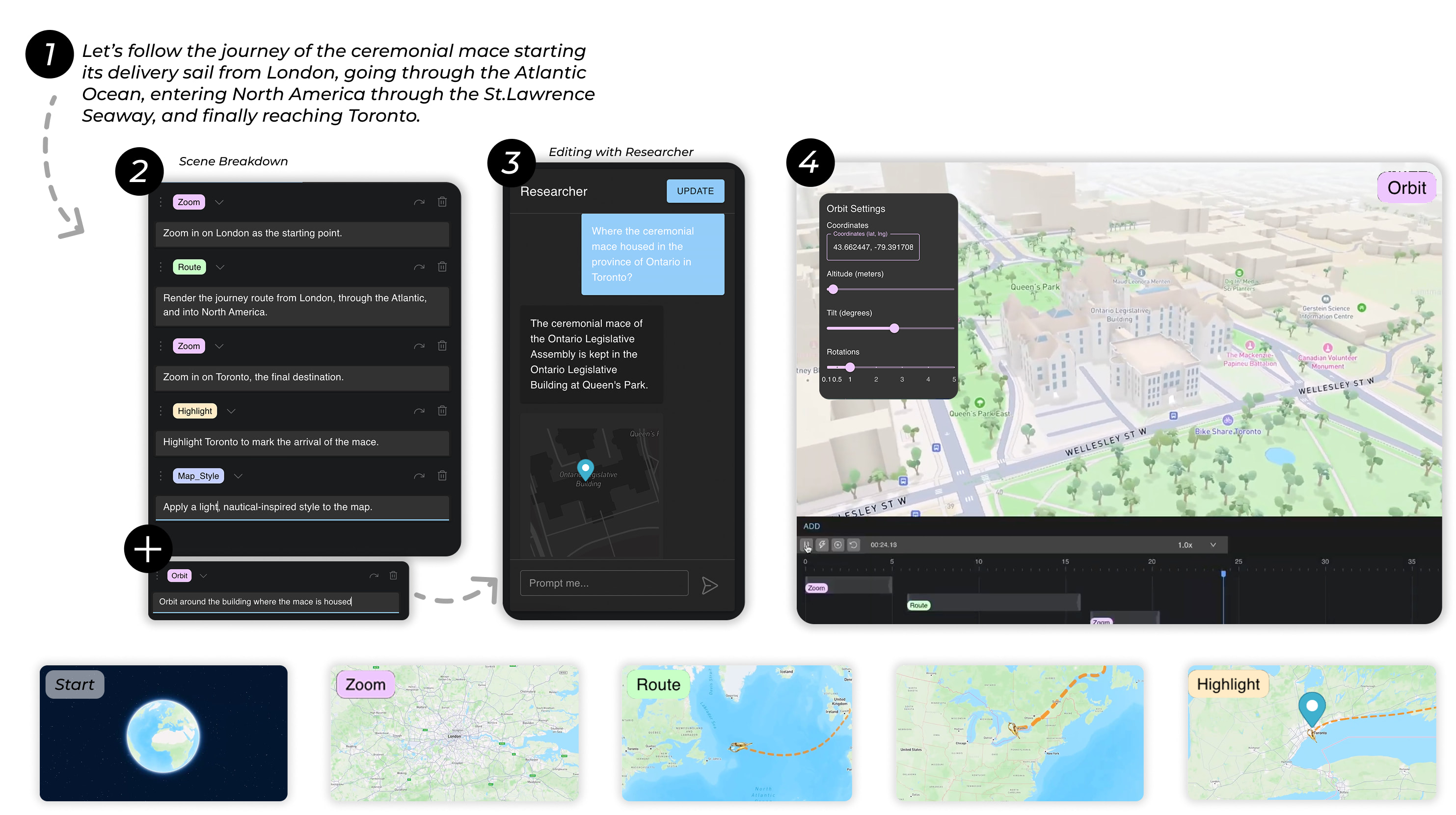} 
  \caption{MapStory System Walkthrough. }
  \label{fig:walkthrough}
\end{figure*}

To establish our map animation primitives, we began by identifying the most common techniques used in existing map animation videos. To this end, we conducted a formative analysis of map-based animations, aiming to pinpoint widely adopted methods. The findings from this study helped shape the system design of our primitive map animation blocks.

\subsection{Method} 
We started from collecting popular map animation videos through YouTube search. Two authors (A1 and A3) began with keyword search, such as \textit{``map animation''}, \textit{``animated maps''}, and more context-specific terms like \textit{``history of [geographic location]''}. Since there is no single dominant keyword to discover these map animation videos, we used YouTube’s recommendation system based on the initial subset of relevant videos to discover additional related videos. We included only videos that employed animation techniques using a map as the primary visual canvas. Static map images or videos without any animated transitions or effects were excluded. Through this process, we curated a total of 200 map animation videos.

Recognizing that each video might feature multiple animation techniques across various scenes, A1 and A3 extracted screenshots from each distinct animation scene, producing a total of 1200 screenshots. These screenshots were then categorized with thematic coding. A1 initiated the coding by organizing the screenshots on a Miro board. After this initial categorization, all authors reviewed the emerging themes and iteratively refined them until reaching a consensus.

\subsection{Results of Map Animation Techniques}
This analysis resulted in the identification of three primary categories of map animation techniques, which we implemented as core components in our system (Figure~\ref{fig:taxonomy}). A complete list of the analyzed videos, along with their corresponding visual overlays, is provided in the supplementary materials.

\subsubsection*{\textbf{Highlights}}
Highlights are a primary method of emphasizing specific geographic areas of interest on a map. They can be implemented as areas, lines, or points:

\noindent
\begin{list}{$\bullet$}{\leftmargin=10pt \itemindent=0pt}
\item \textbf{\textit{Area highlights}} are used to visually emphasize regions or zones by shading, coloring, or bounding the chosen area. This approach effectively focuses the viewer’s attention on a specific region of the map---such as a country, state, or historical territory---by distinguishing it from its surroundings.
\item \textbf{\textit{Line highlights}} outline or trace boundaries and routes. For example, line highlights are often used to represent journeys like troop movements and migration paths. Also, it can show borders, national boundaries, or coordinate references to provide contextual grounding or to show expansion over time.
\item \textbf{\textit{Point highlights}} are small markers, symbols, or icons placed at particular coordinates. These highlights typically indicate important landmarks, cities, or data points, making it clear where the viewer should focus. 
\end{list}

\subsubsection*{\textbf{Camera Controls}}
Camera movement plays a critical role in map-based animation, guiding the viewer's focus and creating dynamic transitions between scenes.

\noindent
\begin{list}{$\bullet$}{\leftmargin=10pt \itemindent=0pt}
\item \textbf{\textit{Zooming}} is the most common camera motion, used to shift the level of detail by moving in and out of specific regions. It helps create narrative pacing and emphasizes points of interest.
\item \textbf{\textit{Translation}} involves shifting or panning the camera’s viewpoint across the map. This technique can guide the audience from one geographic region to another, showcasing transitions between key locations or highlighting spatial relationships.
\item \textbf{\textit{Orbiting}} refers to rotating the camera around a focal point or region on the map. By orbiting, animators can present multiple sides or angles of a location, adding a three-dimensional feel and helping viewers grasp the scope or scale of the geography in question.
\end{list}
These camera movements are frequently orchestrated using keyframes, which define the camera’s position and orientation at specific moments in time, resulting in smooth and deliberate visual transitions.

\subsubsection*{\textbf{Animated Elements}}
Animated elements bring dynamic motion to the map, helping convey movement, changes in territory, or shifting data.

\noindent
\begin{list}{$\bullet$}{\leftmargin=10pt \itemindent=0pt}
\item \textbf{\textit{Route motion}} animates objects, such as ships, planes, and arrows, along paths to illustrate travel, trade routes, invasions, or other directional flows. This technique is especially effective for showing temporal progression or cause-and-effect across space.
\item \textbf{\textit{Spatial transitions}} involve the enlargement or reduction of highlighted regions or shapes over time. This technique is typically used to showcase how areas grow or shrink---such as expanding boundaries in historical conquests, shifting population zones, or changes in environmental conditions.
\item \textbf{\textit{Auxiliary motion}} includes any additional, supportive movement layered onto the map for emphasis or context. Examples include arrows sweeping across a region, clusters of icons like armies or clouds moving in a swarm-like fashion, or scanning beams that pass over key areas. This motion is often employed to guide the viewer’s attention and reinforce important narrative points or data trends.
\end{list}

%% file: 5-system.tex
\section{\system{}}

This section presents \system{}, an interactive tool that generates map-based animations from natural language input. 

\subsection{Overview} 
Built on large language models, \system{} provides the authoring interface that translates text into modular, editable map animations. \system{} takes a text script as input and outputs a sequence of structured animation components, offering flexibility and fine-grained control. \system{} decomposes the input into discrete animation steps by identifying high-level actions, such as zoom, highlight, and route animation, which are represented as editable JSON and later automatically converted into corresponding animations. These modules are visualized on a map canvas and can be interactively refined using a timeline editor and a properties panel. Users can edit the animation modules on a timeline, adjust parameters, and preview changes in real time. To support accurate and fact-grounded animations, users can ask follow up queries and modifications to the system's initial results. \system{} assists animators in resolving vague or ambiguous location references, retrieving precise geospatial data like points, regions, or paths, based on the LLM-powered web search, that is directly integrated into the animation modules.

\subsection{System Walkthrough}
We now walk through an example of how a user can create an animated scene using \system{}, as illustrated in Figure~\ref{fig:walkthrough}.

\subsubsection*{\textbf{Step 1: The User Types a Script into MapStory}}
The user begins by writing, in natural language, a description of the animation scene they wish to create. In Figure~\ref{fig:walkthrough}-1, the user provides the script: \textit{``Let’s follow the journey of the ceremonial mace starting its delivery sail from London, going through the Atlantic Ocean, entering North America through the St. Lawrence Seaway, and finally reaching Toronto.''} The input can be as vague or as detailed as the user prefers. \system{} is designed to interpret a wide range of user intentions from open-ended descriptions. 

\subsubsection*{\textbf{Step 2: The System Breaks Down the Script into Animation Sequences}}
From the initial text input, the system's \textit{\textbf{scene breakdown}} module generates a step-by-step breakdown of the scene into modular map animation primitives. In this example, the system automatically generates the following sequence: \textit{``a zoom to London''}, \textit{``a route animation across the Atlantic''}, \textit{``a zoom to Toronto''}, and \textit{``a highlight on Toronto''}. Each step corresponds to a specific primitive animation block, which can be edited, reordered, added, or removed by the user (Figure~\ref{fig:walkthrough}-2). The user can further refine each block by writing contextual or descriptive text, which the system uses to better understand the animation intention. 

\subsubsection*{\textbf{Step 3: The User Interactively Retrieves Geospatial Map Data}}
Next, the system helps retrieve and process relevant geospatial data via the \textbf{\textit{researcher}} module. This includes fetching GeoJSON coordinates for locations, areas, or routes mentioned in the script. The system first automatically fetches the relevant information for each block in the scene breakdown. However, the user can follow up with additional queries or modifications. For example, when the user asks, \textit{``Where is the ceremonial mace housed in the province of Ontario in Toronto?''}, the system identifies the Ontario Legislative Building as the destination and extracts its location for use in the animation  (Figure~\ref{fig:walkthrough}-2). The system can also pull in additional metadata, such as historical context, the year a building was constructed, or statistics like the number of medals won by a country, allowing users to focus on their story.

\subsubsection*{\textbf{Step 4: The User Modifies Parameters, Adjusts Styles, and Controls Timing of Each Animation Module}}
With the animation structure and map data in place, users can fine-tune the look and feel of each element using the properties panel. For example, they can add images, label annotations, or change the style to enhance the animation. Each module’s visual properties, such as color, transparency, labels, and styling, can be customized. For instance, a user might highlight a country in a vivid color, adjust the opacity of a route, or change the map style to match the narrative tone.

Finally, users can interactively control the timing of each animation module by specifying when it appears and how long it plays, through the timeline sequencer. This allows for fine-grained control over pacing and sequencing in real time. By iteratively refining timing and transitions, users can craft a cohesive and compelling animated story that aligns with their intended narrative. The final result is a text-driven animation that seamlessly integrates geospatial data, visual styles, and temporal coordination.

%% file: 6-implementation.tex
\section{Implementation}
Our system is implemented using React.js and TypeScript, composed of three main components: 1) a base map canvas built with Mapbox, 2) a custom timeline editor and animation sequencer, and 3) a text-driven scene breakdown and researcher interfaces powered by our large language model architecture. We use OpenAI’s \textit{o1} as the primary multimodal language model to process user input and assist with animation generation. For geospatial queries, the system uses Perplexity's \textit{sonar-pro} model to retrieve and integrate geospatial data, which has access to the web. We additionally query GeoJSON data from OpenStreetMap via the Nominatim API~\footnote{\url{https://nominatim.openstreetmap.org/ui/search.html}}, enabling location lookup and metadata retrieval based on natural language queries. For map animation rendering, we use the Mapbox API within the React.js framework. The system can also dynamically change the map's appearance by selecting from a variety of styles supported by Mapbox~\footnote{\url{https://docs.mapbox.com/api/maps/styles/}}.

\begin{figure}[h!]
\centering
\includegraphics[width=0.32\linewidth]{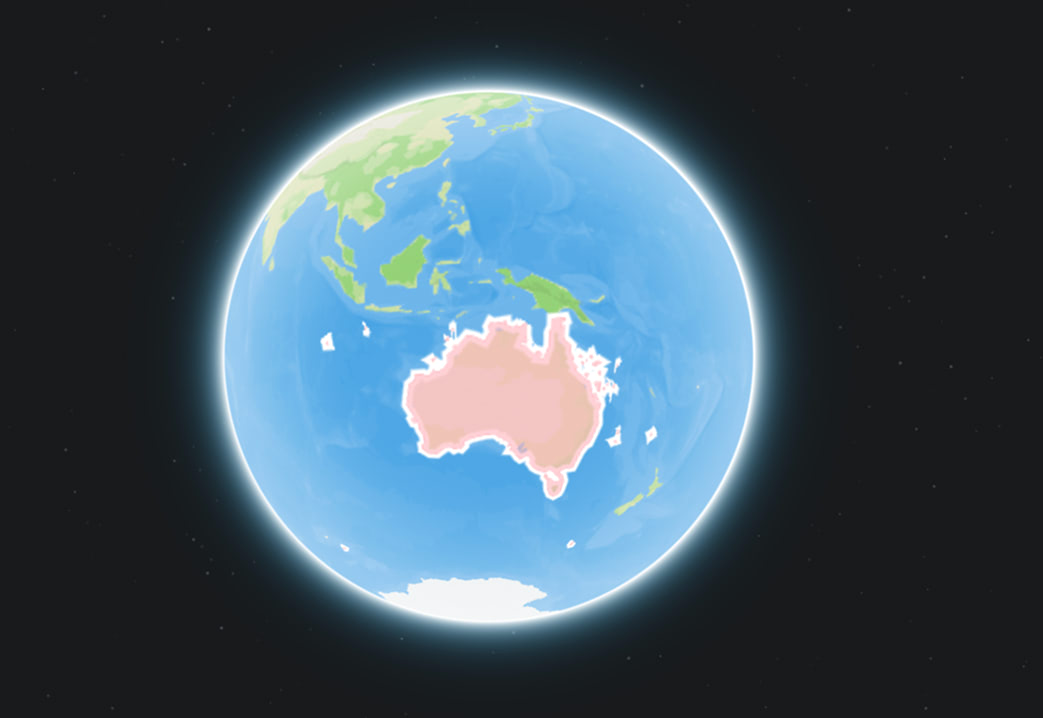}
\includegraphics[width=0.32\linewidth]{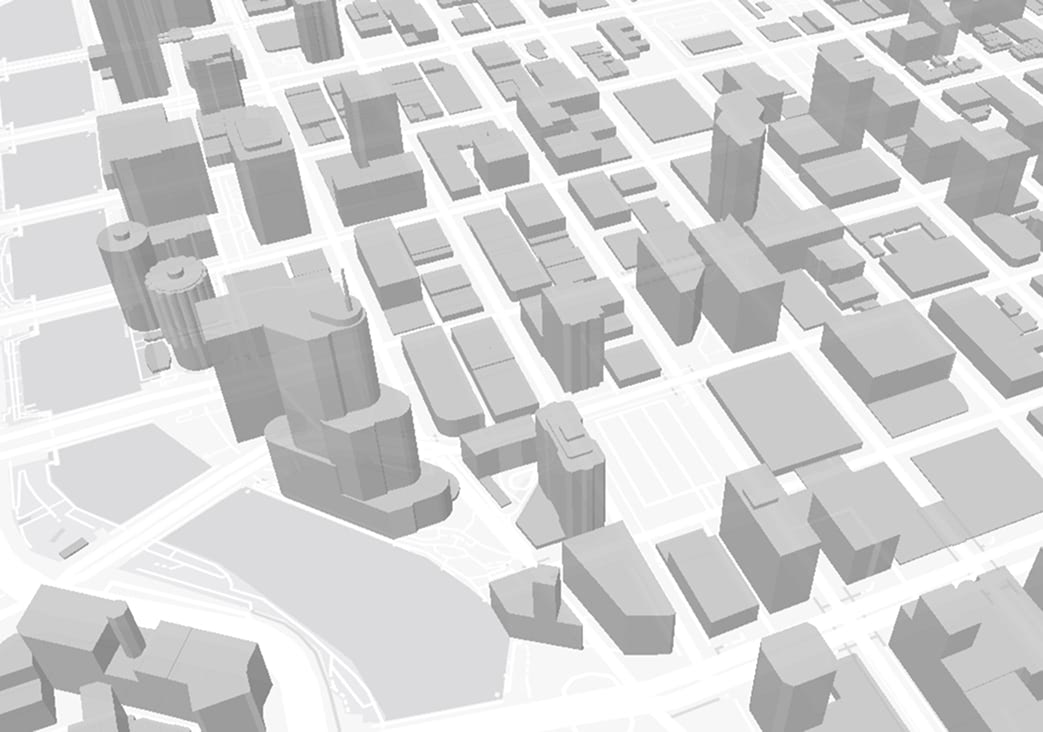}
\includegraphics[width=0.32\linewidth]{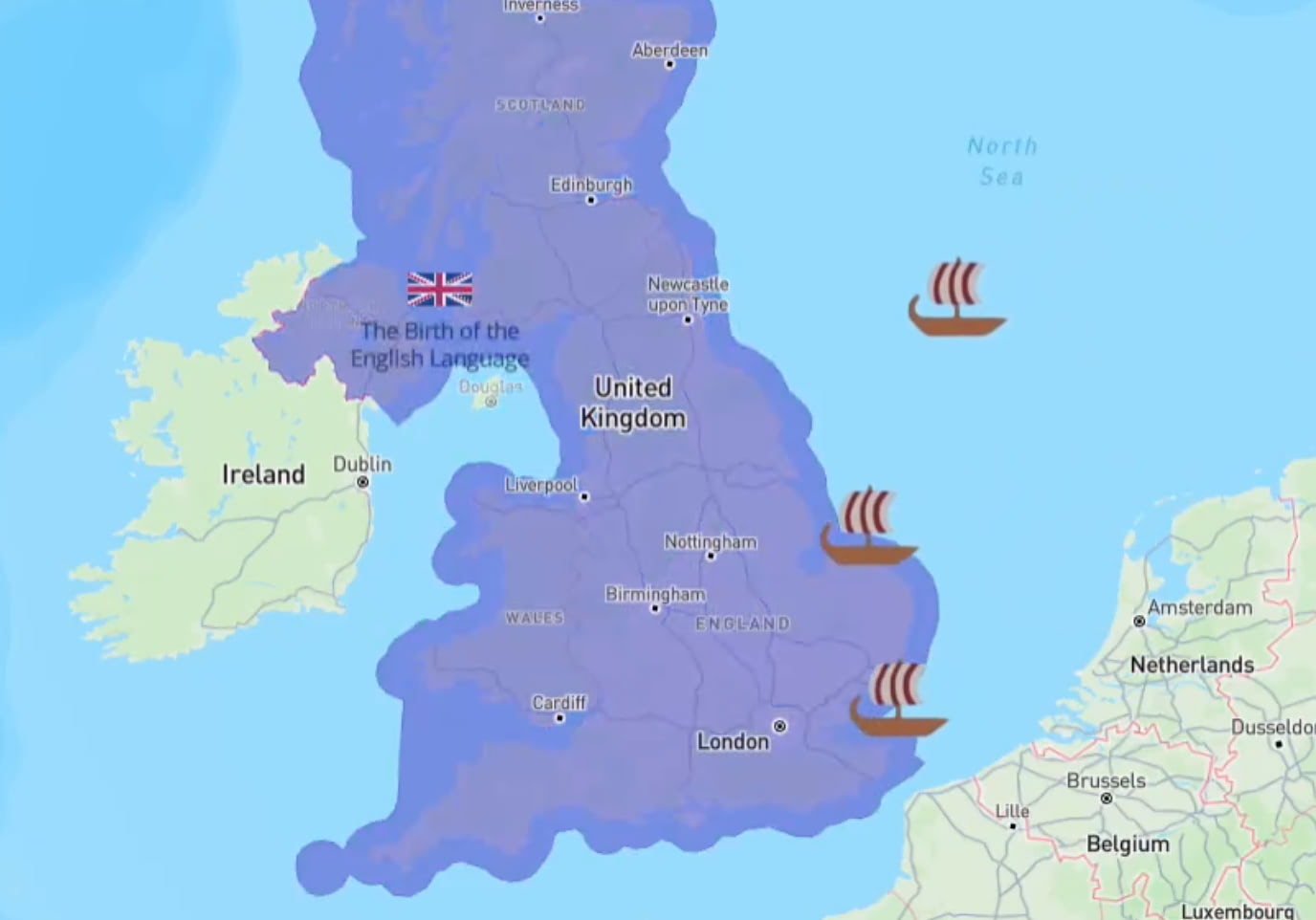}
\caption{Map Stylization}
\label{fig:mapstyle}
\end{figure}

\subsection{Animation Sequencer}

The animation sequencer, which is visually represented as a timeline is implemented in React.js in the browser environment using Mapbox API as a base canvas. We use Fabric.js to render overlays on top of this canvas. The engine comprises of a time interval which runs every second when the user clicks the play button. Each map animation primitive in the system is represented as with a structured JSON block with a \textit{block\_name}, \textit{start\_time}, \textit{end\_time}, and \textit{block\_args}, where the block arguments are specific to each block type. These blocks are typically created by the LLM agent but can be manually added by the user as well. When the animation starts, the renderer periodically checks if the time interval matches the start time of any block. If it does, a block specific function is called to start the animation sequence of that block. These block specific functions  animate and visualize objects on the Mapbox canvas implemented as a fabric.js overlay. The block arguments with start and end times are passed into the block function, which then performs the animation action rendered on screen.

\subsection{Supported Primitive Map Animation Blocks}
In this section, we describe how our system implements the core map animation primitives. For each primitive block, the choice of supported animation is informed by our taxonomy analysis, and each category corresponds to the descriptions provided in Section 4. These modules serve as the fundamental \textit{animation primitives} in our system. 

\subsubsection*{\textbf{Highlights}}
The Highlight block supports three types of visual emphasis: \textit{Area}, \textit{Line}, and \textit{Point}.

\noindent
\begin{list}{$\bullet$}{\leftmargin=10pt \itemindent=0pt}
\item \textbf{\textit{Area}}: Renders a GeoJSON polygon on the map canvas at a specified time interval. The polygon data is fetched from the Nominatim API or generated by the LLM.
\item \textbf{\textit{Line}}: Uses an array of latitude-longitude coordinates provided by the LLM to render a polyline on the map.
\item \textbf{\textit{Point}}: Plots a map marker at the specified coordinates.
\end{list}
Users can freely attach text or images to highlights of any type, offering a flexible way to annotate or enhance the visual presentation of geographic features.

\begin{figure}[h!]
\centering
\includegraphics[width=0.32\linewidth]{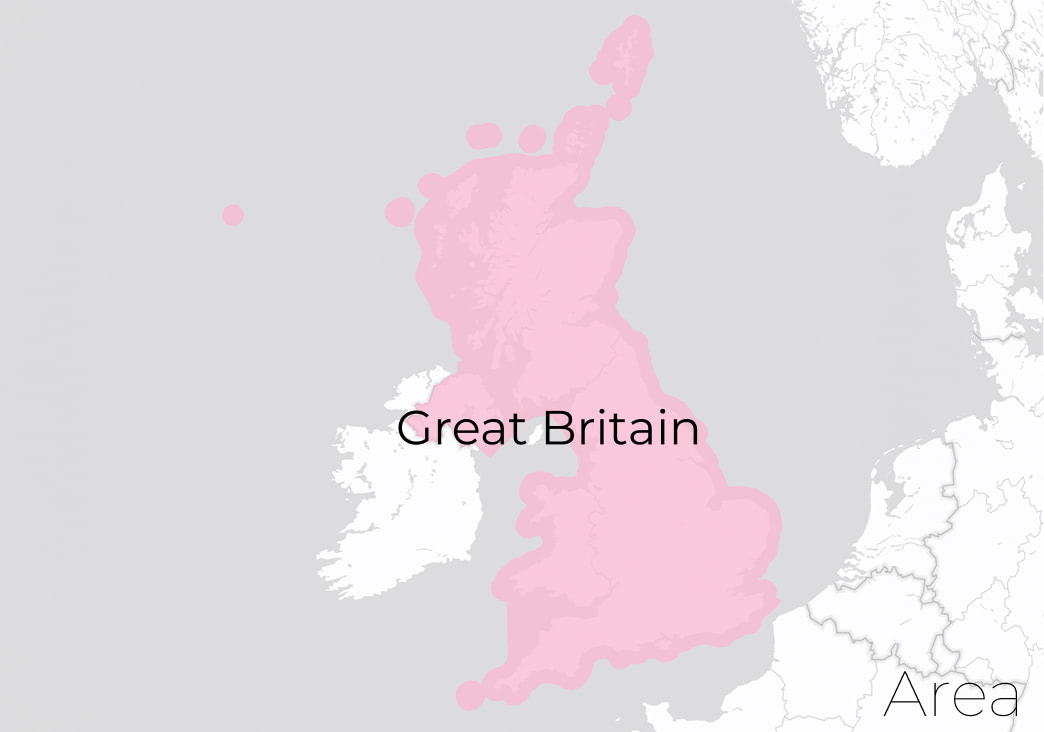}
\includegraphics[width=0.32\linewidth]{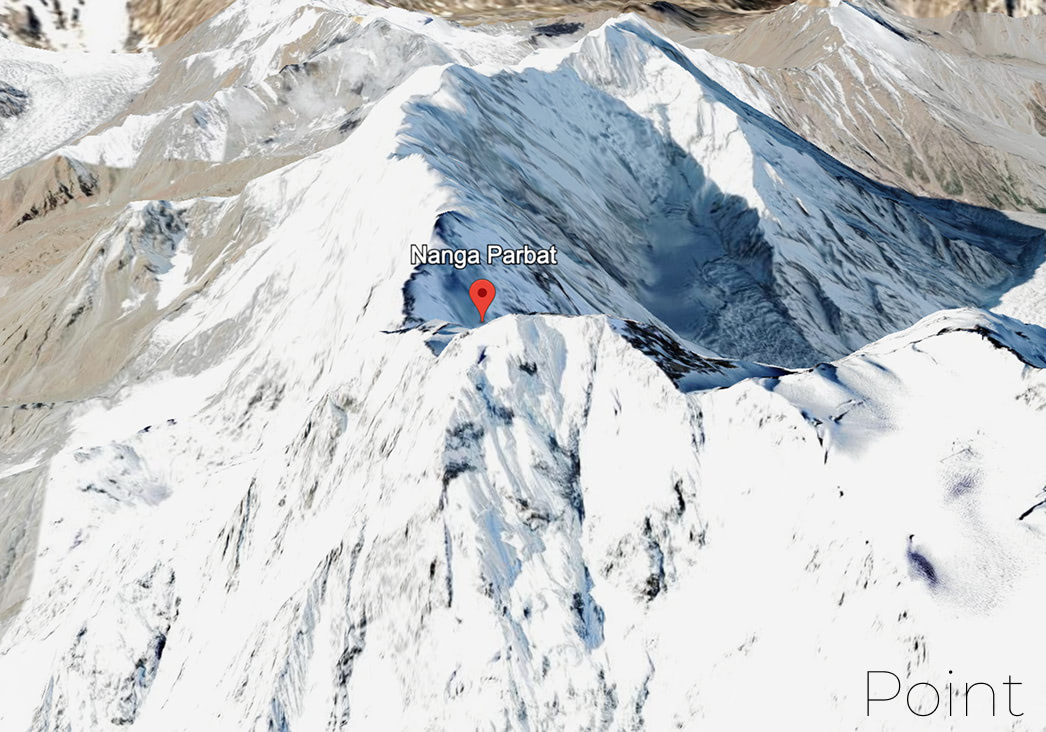}
\includegraphics[width=0.32\linewidth]{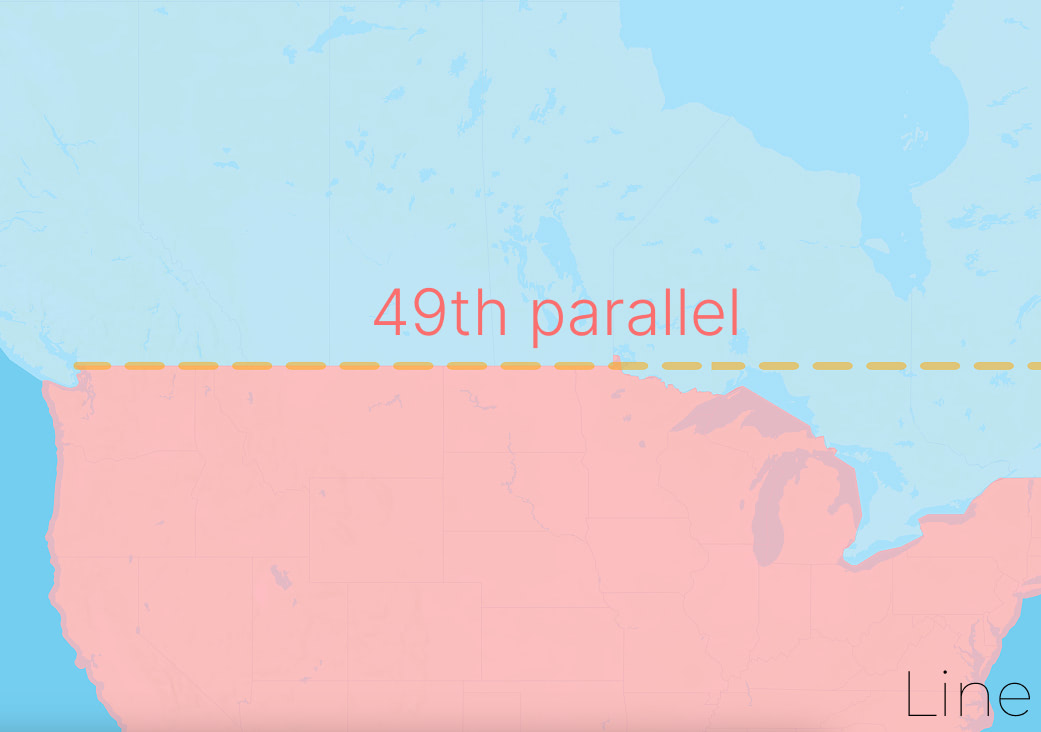}
\caption{Our system supports three types of highlights 1) area, 2) point, and 3) line.}
\label{fig:highlight}
\end{figure}

\subsubsection*{\textbf{Camera Controls}}
Camera movements are handled by animating the viewpoint in Mapbox:

\noindent
\begin{list}{$\bullet$}{\leftmargin=10pt \itemindent=0pt}
\item \textbf{\textit{Zoom}}: Employs the built-in FlyTo function to smoothly zoom into a specified location.
\item \textbf{\textit{Translate}}: Interpolates the camera’s target position from start to end coordinates, shifting the focal point without changing the zoom level.
\item \textbf{\textit{Orbit}}: Moves the camera along a circular path using incremental angular steps every frame, creating a revolving view of a given focal point.
\end{list}

\begin{figure}[h!]
\centering
\includegraphics[width=0.32\linewidth]{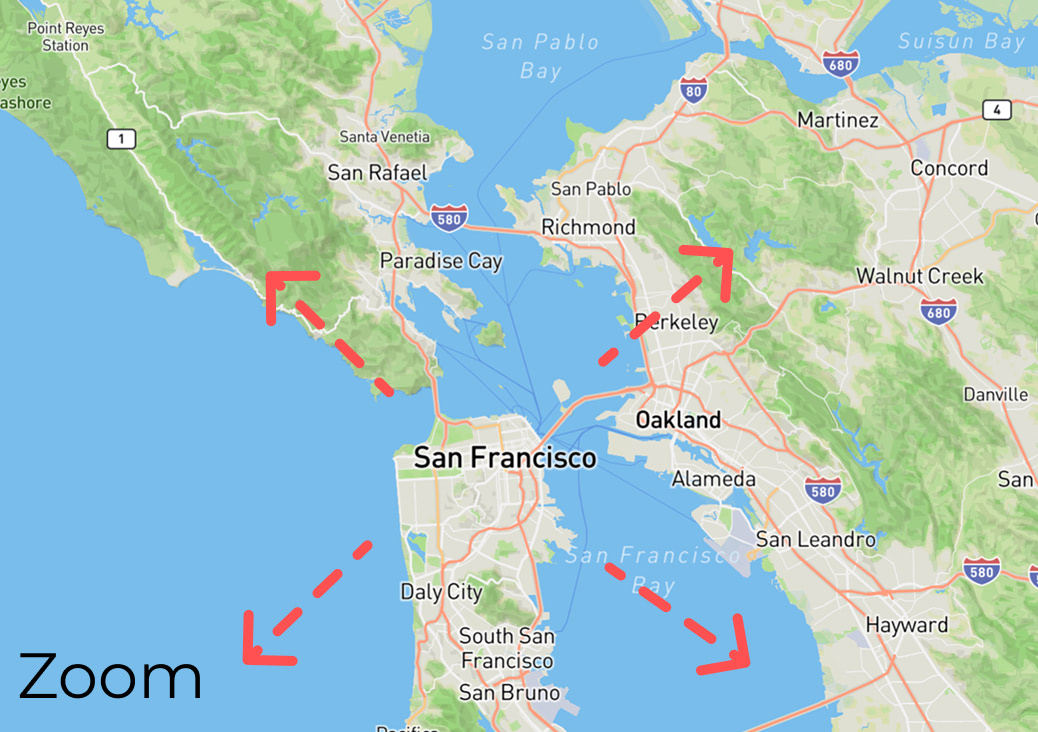}
\includegraphics[width=0.32\linewidth]{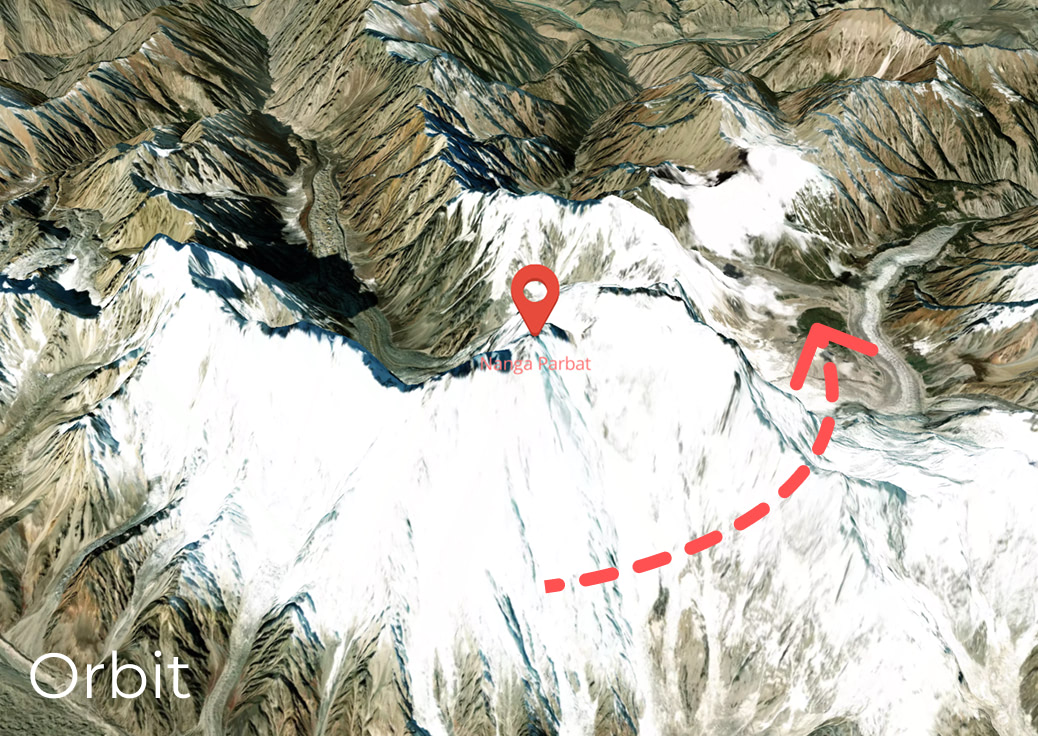}
\includegraphics[width=0.32\linewidth]{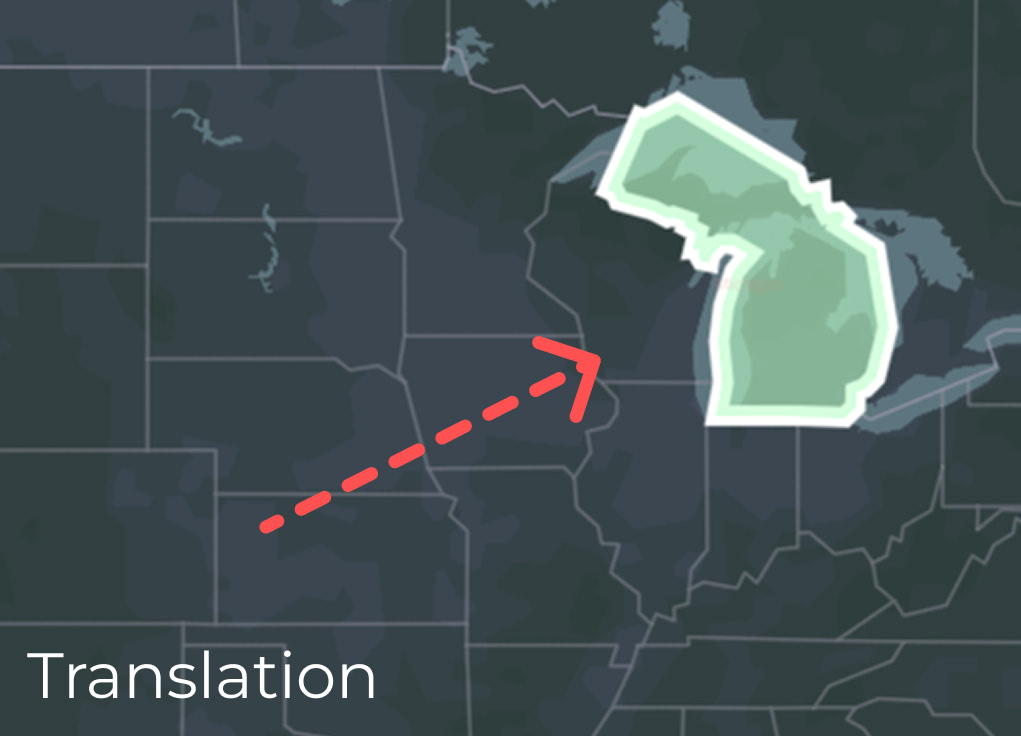}
\caption{Camera controls: 1) zoom, 2) orbit, and 3) translation.}
\label{fig:cameracontrols}
\end{figure}

\subsubsection*{\textbf{Animated Elements}}
To create dynamic motion atop the map canvas, our system employs a custom animator built using browser intervals and Fabric.js\footnote{\url{https://fabricjs.com/}}. By defining sprite positions over time in latitude–longitude format, we can display animation frames that smoothly transition across the map.

\noindent
\begin{list}{$\bullet$}{\leftmargin=10pt \itemindent=0pt}
\item \textbf{\textit{Animated Routes}}: Drawn via the Mapbox drawing API, routes are specified as an array of latitudes and longitudes, generated by the LLM (or modified by the animator). Once the path is rendered, user-uploaded sprites can be animated along that route. These sprites remain fully editable, and their position, style, or any other attribute can be adjusted.
\item \textbf{\textit{Spatial Transitions}}: Implemented as polygon shape transformations or translations. For example, to show a transition between the polygons for North Dakota and a merged North–South Dakota shape, we interpolate these two polygons over a specified time interval with Flubber \footnote{\url{https://www.npmjs.com/package/flubber/v/0.1.0}}, creating a morphing animation. Simple translations move the polygon from one location to another.
\item \textbf{\textit{Auxiliary Motion}}: Provides secondary movements such as looping animations across a specified coordinate range on the map. Users can upload sprites, which the system duplicates into a cluster. The LLM determines the motion range and cluster count, enabling versatile and visually engaging secondary animations.
\end{list}

\begin{figure}[h!]
\centering
\includegraphics[width=0.32\linewidth]{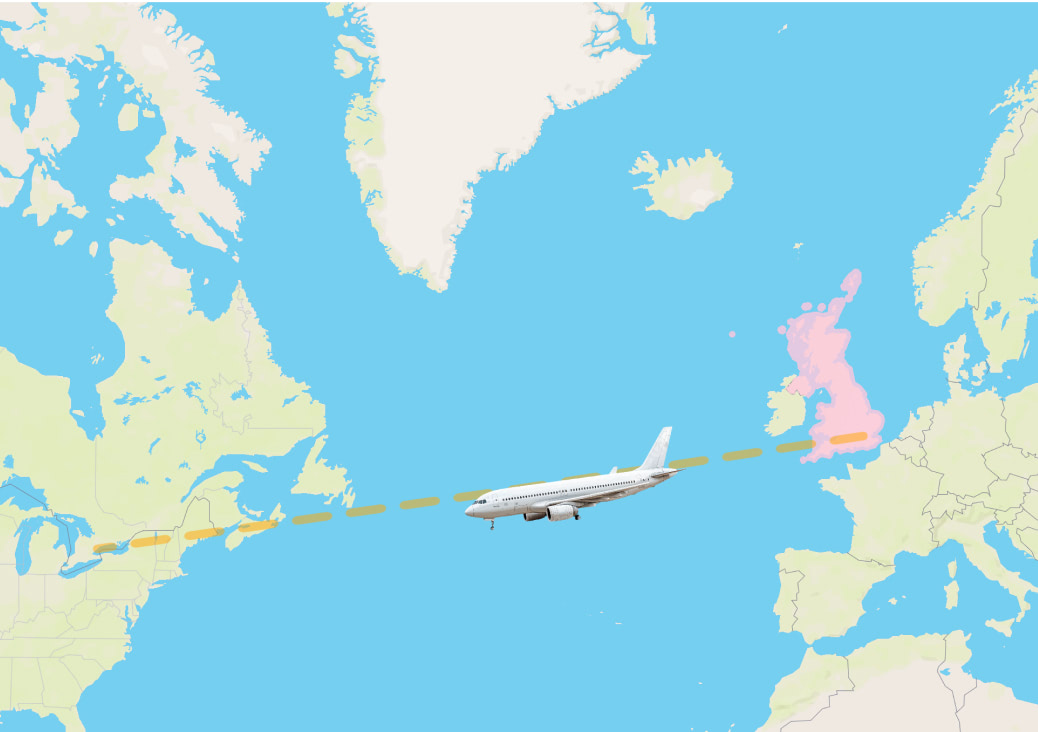}
\includegraphics[width=0.32\linewidth]{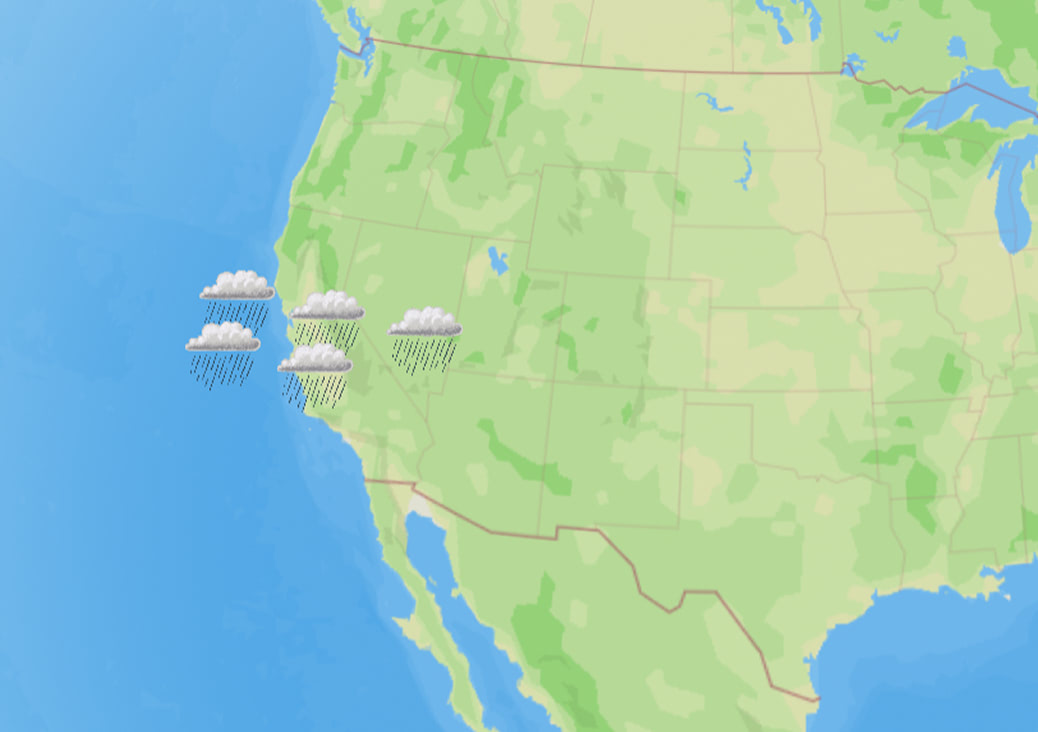}
\includegraphics[width=0.32\linewidth]{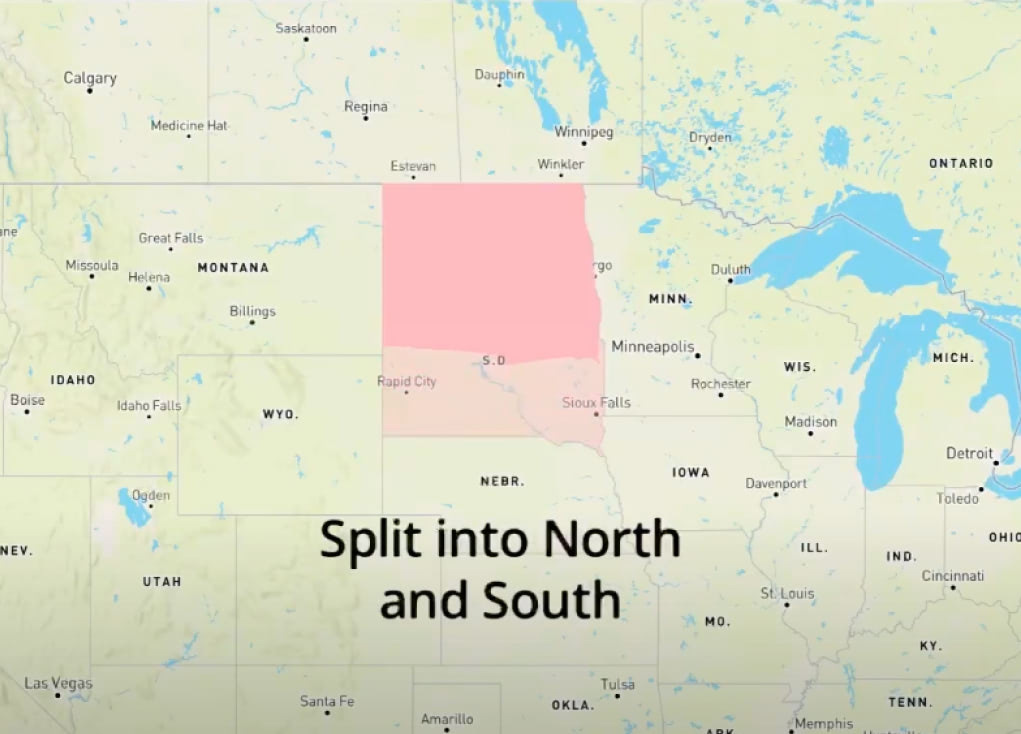}
\caption{Animated elements: 1) routes, 2) auxiliary motion, and 3) spatial transitions.}
\label{fig:animatedelements}
\end{figure}

\subsection{LLM Architecture}

\begin{figure*}
\noindent
\includegraphics[width=1\textwidth]{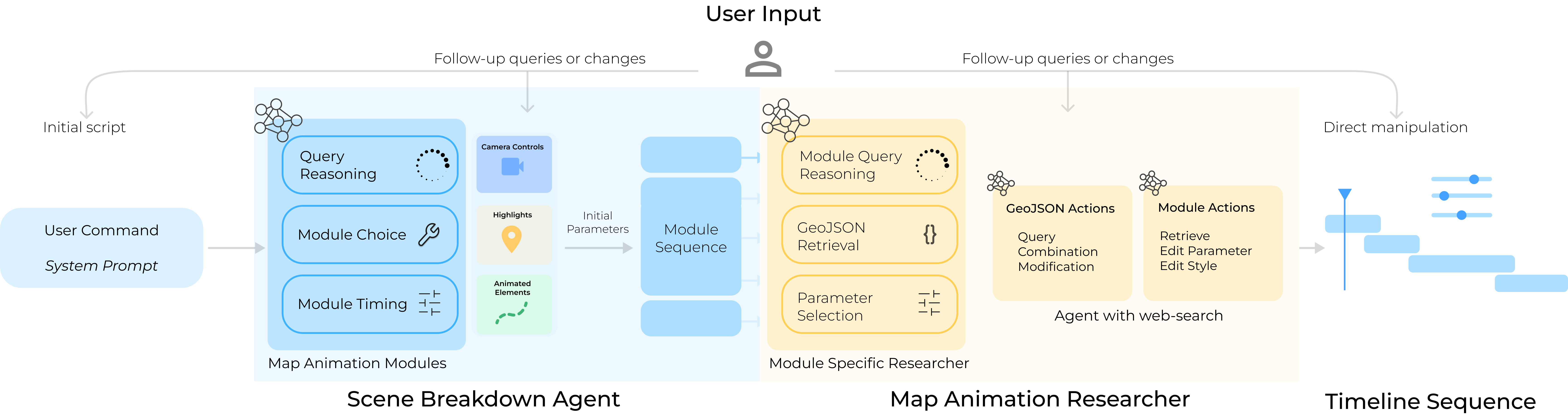} 
\caption{MapStory LLM Architecture}
\label{fig:architecture}
\end{figure*}

\system{}'s agent architecture follows a human-in-the-loop approach to creating, verifying, and querying information for map animations. We chose to separate our architecture into an agentic workflow consisting of multiple calls as LLMs perform better when broken down into modular task due to token limitations, continuity issues or clarity \cite{wangprompting}. We chose the two main tasks of creating map animations as found in our formative explorations: an animation planning stage and a research stage. Our architecture features two primary LLM agents: a scene breakdown agent and a map animation researcher agent. We use the \textit{o1} model for scene breakdown tasks, while Perplexity’s \textit{sonar-pro} model serves as the research agent, helping with web-queried outputs. The scene breakdown agent is responsible for converting user input into a structured JSON-based scene layout with key animation blocks in sequence. The researcher agent then examines the relevant modules from that breakdown, performs chain-of-thought reasoning \cite{wei2022chain} to validate their parameters, and finally invokes the appropriate function call with the confirmed arguments. Below, we detail both agent roles.

\subsection*{\textbf{Scene Breakdown Agent}}

The scene breakdown agent is designed to create step-by-step editable animation guides \textbf{(D1)}, which follow an animator’s traditional approach of creating animation guides from their scripts. The scene breakdown agent is implemented as a single function call and can be re-prompted with additional queries by the user interactively.  Given a user $script$, the agent returns a sequence of primitive map animations with each item containing a \textit{module\_name}, \textit{timeline\_start}, \textit{timeline\_end}, \textit{short\_description}, \textit{long\_description} and \textit{initial\_parameters} (module specific). We describe the steps the agent takes to achieve this below.



\subsubsection*{\textbf{Query Reasoning}}

Given an initial script, the scene breakdown agent is first instructed to first perform query reasoning to infer the intent of the user and instructed to simplify the text into a list of requirements. The scene breakdown agent follows a planning approach where a task is divided into smaller manageable sub-tasks \cite{wang2023plan} while internally conducting CoT reasoning \cite{wei2022chain}.

\subsubsection*{\textbf{Module Choice}}
The agent is then instructed to deconstruct the initial free-form user query into actionable animation guides in plain text.  The prompt includes descriptions of each animation block <\textit{primitive\_descriptions}> supported by our system within three categories: \textit{highlight}, \textit{camera control}, and \textit{animated element} blocks. The prompt also includes the following heuristic consideration: Always precede highlight or animated element blocks with a \textit{zoom} or \textit{translate} blocks to set the focus on the area.

\subsubsection*{\textbf{Module Timing}}
The agent is instructed to schedule each module in strictly consecutive, discrete, non‑overlapping intervals with no gaps with the following heuristic considerations: 1) avoid overlapping camera control blocks; 2) time modules so that, at any given camera focus, all active modules operate within the same visible region; and 3) the mapstyle module must not overlap with any other module and, unless otherwise specified, should span the entire sequence.

\subsubsection*{\textbf{Initial Parameter Tuning}}

For each animation primitive module, the agent is instructed to produce both a short and a long description for each block. The \textit{short\_description} is one sentence describing the module and its main parameter. For the \textit{long\_description}, the agent sets initial parameters for the animation block, such as the latitude and longitude of the location, the initial query for the Nominatim API, or an initial list of coordinates for a route. This provides additional and specific context for the researcher to perform reasoning, querying, and verification.


\subsection*{\textbf{Map Animation Researcher Agent}}

Once the scene breakdown agent has drafted an initial set of animation guides and set initial parameters, the researcher agent parses through each item with its unique user-specified and system-generated context to find or generate appropriate GeoJSON and map-specific coordinates. The researcher is designed to reason about the user’s intent and extract the relevant information from the web or other sources \textbf{(D2)}. We use Perplexity’s \textit{sonar-pro} with web-search and reasoning capability to query, search, and fetch relevant GeoJSON data to satisfy the user’s specific request.

A dedicated research agent is assigned to each animation module. The module specific researcher agent validates initial parameters from the scene breakdown agent and each research agent’s system prompt is tailored to a specific animation block. Additionally, each research agent can only access one tool: a function call adhering to the OpenAI Function Calling Protocol \cite{OpenAI_FunctionCalling_2023}. This modular structure not only streamlines the system’s architecture but also minimizes context length, given the inverse relationship between context size and instruction-following accuracy~\cite{liu2023lostmiddlelanguagemodels}. 

We break down the agent prompt into the following parts:

\subsubsection*{\textbf{Query Reasoning}}

Given a scene breakdown object with user query, long and short description, initial parameters, and overall scene breakdown, the researcher module is first instructed to reason about user intent, disambiguating queries into map-compatible or queryable terms. For example, when asked to highlight “\textit{Andhra Pradesh before 2014},” the researcher is instructed to first verify if this is a Nominatim-compatible query. In this case, it is not, as the user is asking for region boundaries in the past. It disambiguates ambiguous terms and flags any request that cannot be satisfied by a direct Nominatim lookup (e.g., “Andhra Pradesh before 2014,” which requires historical rather than current boundaries). If not, the agent deconstructs the query into a list of \textit{n} Nominatim compatible strings which best match the query description. This will later be fetched from the Nominatim call.


\subsubsection*{\textbf{GeoJSON Retrival}}

Once query reasoning is complete, the agent moves on to the GeoJSON retrieval step. We use the OpenStreetMap Nominatim API to query and retrieve various types of GeoJSONs, like polygons, multipolygons, lines, or points. The arguments taken by the API are a compatible \textit{query\_string} and optional arguments like \textit{polygon\_type}, \textit{country}, \textit{viewbox} etc. A full list of compatible parameters for the API can be found here \footnote{\url{https://nominatim.org/release-docs/latest/api/Search/}} For example, the “Avon River” can be found in both the UK and Canada. We can construct a query to disambiguate between the two by replacing the bounding coordinates \href{https://nominatim.openstreetmap.org/search?q=River+Avon&format=geojson&polygon_geojson=1&countrycodes=GB&viewbox=-2.8,51.3,-1.5,51.6&bounded=1}{(\textit{Example query with the UK as filter})}. The researcher agent is then instructed to choose an appropriate \textit{GeoJSON retrival action}.


\subsubsection{\textbf{GeoJSON Actions}}

\change{We define actions as a supplementary query to a specialized agent which performs a modify action on the fetched GeoJSON for a given query. Our system supports the following modifications on fetched GeoJSONs:}

\begin{list}{$\bullet$}{\leftmargin=10pt \itemindent=0pt}\item \textit{\textbf{Query}}: If a query is determined to simply be Nominatim-compatible, the agent constructs a compatible query and fetches the GeoJSON. For example, querying “\textit{Andhra Pradesh}” results in a polygon of the state in India in the present day.

\item \textit{\textbf{Combination}}: The combination action performs basic set operations (\textit{Union, Difference, and Intersection}) on the fetched GeoJSON polygons to form the desired boundary. The agent first returns a string array of all Nominatim‑compatible queries that satisfy a request. The agent is instructed to select the operation that minimizes the number of Nominatim queries.  For example, for “Andhra Pradesh in 2014,” it retrieves the boundaries of the constituent regions \textit{“Telangana, India,” “Andhra Pradesh, India”} and applies a \textit{union}. In another case, it may choose \textit{intersection}; for instance, to fetch only the Bow River segment within Calgary, the agent intersects the queried river and city geometries to return the correct GeoJSON.

\item \textit{\textbf{Modification}}: The modification action enables the LLM to directly change the GeoJSON based on the query intention for arbitrary stylistic edits.
    \begin{itemize}
      \item \textit{\textbf{Reduction}}: Removes unwanted sub‑features via LLM‑driven JSON edits. For example, to visualize evolving historical borders not present‑day, the agent deletes arbitrary shapes as requested by the user and emits the updated boundary without any set operations.
      \item \textit{\textbf{Generation}}: For simple requests that are not queryable, such as representative flight or sea routes (which the agent does not have a database to query), the agent generates an estimated GeoJSON with key waypoints.
    \end{itemize}

\end{list}

\subsubsection*{\textbf{Parameter Selection}}
The agent is then instructed to select relevant module parameters. For example, if a highlight module requires text to display the number of gold medals won by a country or if a route color needs to be changed to contrast with the map style. Finally, the response (which relies on the function calling API) is parsed into a compatible JSON output with the animation functions available in the system. This is then rendered as an editable timeline. 
The agent can perform \textbf{module actions} to change the module animation in three ways: 1) It can retrieve or change relevant geospatial information, 2) It can edit module-specific parameters (such as zoom levels) based on the narrative requirements, and 3) It can edit the style of the module based on the script provided by the user. The agent chooses an appropriate action based on the user's request.

\subsubsection*{\textbf{Human-in-the-Loop (HITL) Editing}}
Once the agent generates an initial scene breakdown, the user can edit, rearrange, or delete any part and regenerate a scene breakdown. For example, the user can change the order the blocks appear in, add or delete blocks or edit short and long descriptions of individual map animation blocks to give additional context. Similarly, by accessing the researcher tab when a specific item in the timeline or scene breakdown is selected, the user can communicate with the researcher through a chat interface and ask for specific questions or modifications (\textbf{D3}). The user can iteratively refine the prototypes and follow up with changes via direct manipulation or natural conversation.

%% file: 7-user-study.tex
\section{Technical Evaluation}

To evaluate our system's performance, we conducted a technical evaluation.

\subsection{Method}

\subsubsection*{\textbf{Dataset}}
We used 20 map-based storytelling prompts generated by GPT 4.5 to evaluate our system. These prompts were generated using 5 user-generated prompts as input, with additional context about our system
such as \textit{``In 1492, Christopher Columbus set sail from Palos, Spain, and after weeks at sea, he arrived in the Bahamas, forever altering world history.''}. We then tested these 20 prompts with four different LLM models: 1) GPT-4.5, 2) o1, 3) GPT-4o, and 4) GPT-3.5 to compare the performance. We did not include a baseline because we simply tested our system with text-to-video generation with Sora but with a simple test of 5 prompts, we immediately realized that this approach failed completely, (e.g. generate completely wrong video, generate completely inaccurate information, etc). Therefore, we evaluated our system across different LLM models.



\subsubsection*{\textbf{Procedure}}

We entered each prompt into the Scene Breakdown agent and recorded both the generated modules and the time it took to generate them. After that, we called the Researcher agent to process the modules, with its execution time also measured. After the Researcher completed processing, we added the modules to the timeline and evaluate them. We then graded each module on its accuracy, with criteria varying depending on the module type. 
Camera control modules must have 1) the correct location, and 2) the correct zoom level. Highlight modules must have accurate boundaries, points and lines. Animated element modules must be within 1 km of the real routes, as well as being physically plausible (e.g. train routes musn't go through bodies of water). We evaluated accuracy based on our judgment as no standardized or objective ground truth was available for these types of outputs.
\begin{figure}[h!]
\centering
 \includegraphics[width=\linewidth]{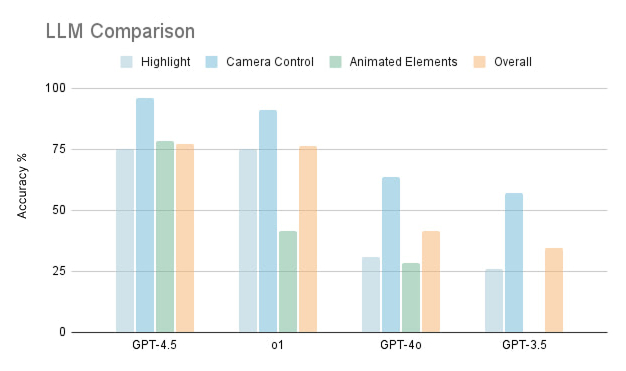}

\caption{Results of accuracy across four LLM models.}
\label{fig:performance01}
\end{figure}


\subsection{Results}


\begin{figure}[h!]
\includegraphics[width=\linewidth]{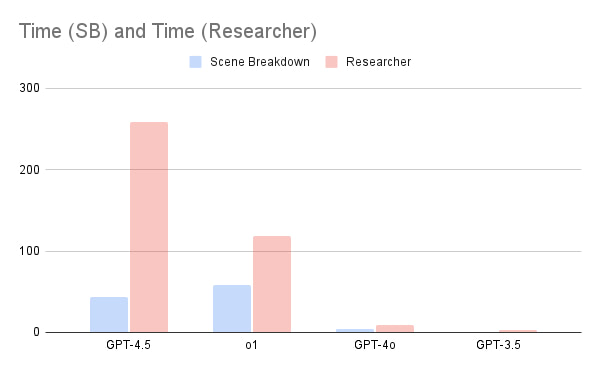}

\caption{Results of processing time (in seconds) across four LLM models.}
\label{fig:time_result}
\end{figure}

Figure~\ref{fig:performance01} presents the accuracy results across the four evaluated models. As expected, we observed a clear improvement in system performance with each successive model. For context, GPT-3.5 is the oldest model, followed by GPT-4o, then \textit{o1}, then GPT-4.5. These results suggest that the capabilities of the underlying LLM play a substantial role in determining the overall performance of our system. As more sophisticated models become available, we expect corresponding improvements in our system’s capabilities.

Figure \ref{fig:time_result} shows the average processing time (in seconds) for each model used in our system. As expected, GPT-4.5 took the longest to produce an output, followed by \textit{o1}, 4o, and 3.5 respectively. Interestingly, GPT-4.5 took less time to produce an output for the scene breakdown, this is presumably due to GPT-4.5 not being a reasoning model. \change{As the efficacy of the models increases so does the time. Failure cases can be attributed to the wrong or inaccurate geocoding by the LLM or complex and historical region queries not available for the LLM to fetch from the databases (even with addition, reduction functions). For example, a test prompt in our technical evaluation involved a path through space to the Moon, which we do not support. Moreover, we observed worse performance when the query involved historical data that needed to be mapped to present‑day, queryable geography. For example, when visualizing the Roman Empire, the system failed to cover important provinces or returned only small fragments of its territory (which corresponds to present day) rather than the full extent required.}

\subsubsection{System Bottlenecks}
\change{The primary bottleneck of the system is the Scene Breakdown Agent. In a single shot (single user prompt), the number of animation modules that can be generated is bound by this agent's token limit. At the time of testing, the GPT-4o output token limit is 16k \cite{dawson2024gpt4o}. We measured the average token size for one animation module as 300. Route modules require the most tokens, and zoom modules require the least. The theoretical limit of our system then becomes ~54 animation blocks for a single user input. }

\section{User Evaluation}

To understand how \system{} supports novice users in creating map animations, we conducted a usability study with twelve participants. We aimed to evaluate the tool’s expressiveness, learnability, support for creative exploration, and factual correctness.

\subsection{Method}

\subsubsection*{\textbf{Participants}}
We recruited 12 participants (5 male, 7 female; ages 21–30, mean=25) via university mailing lists and snowballing sampling. All participants had minimal or no prior experience with animation tools. 

\subsubsection*{\textbf{Tasks}}
Participants were asked to complete two animation tasks:
\begin{list}{$\bullet$}{\leftmargin=10pt \itemindent=0pt}
\item \textbf{\textit{Task A (Guided)}}: Create a short animation based on a pre-written script about the migration of monarch butterflies from Canada to Mexico.
\item \textbf{\textit{Task B (Open-ended)}}: Create an animation based on a self-authored narrative (e.g., historical event, personal story, or travel journey).
\end{list}
\subsubsection*{\textbf{Procedure}}
Each session lasted approximately 45 minutes and was conducted in person. Participants received a compensation of \$15 USD.
Before starting the tasks, participants were given a brief tutorial explaining the study's purpose and the functionality of our system. They were encouraged to ask questions during the tutorial and Task A to ensure a clear understanding. However, to minimize external influence on their creativity, participants were not allowed to ask questions during Task B.
For each task, participants were allowed to work on it until they were satisfied and were encouraged to think aloud during the study.
To evaluate system accuracy, the experimenter recorded the number of animation modules created during each task. Afterward, participants were asked which modules they believed accurately reflected their prompts, allowing us to compute a subjective accuracy rate.
After completing each task, participants filled out a questionnaire covering three aspects: 1) system usability, based on the System Usability Scale (SUS)~\cite{brooke1996sus}, 2) creativity support, based on the Creativity Support Index (CSI)~\cite{cherry2014quantifying}, and 3) usefulness for each system feature, based on our 7-point Likert scale questionnaire. After both tasks, we conducted a semi-structured interviews to gain in-depth feedback on their experience. \change{We include some example videos generated by participants in the video figure.}

\subsection{Results}

\begin{figure}[h!]
\centering
\includegraphics[width=\linewidth]{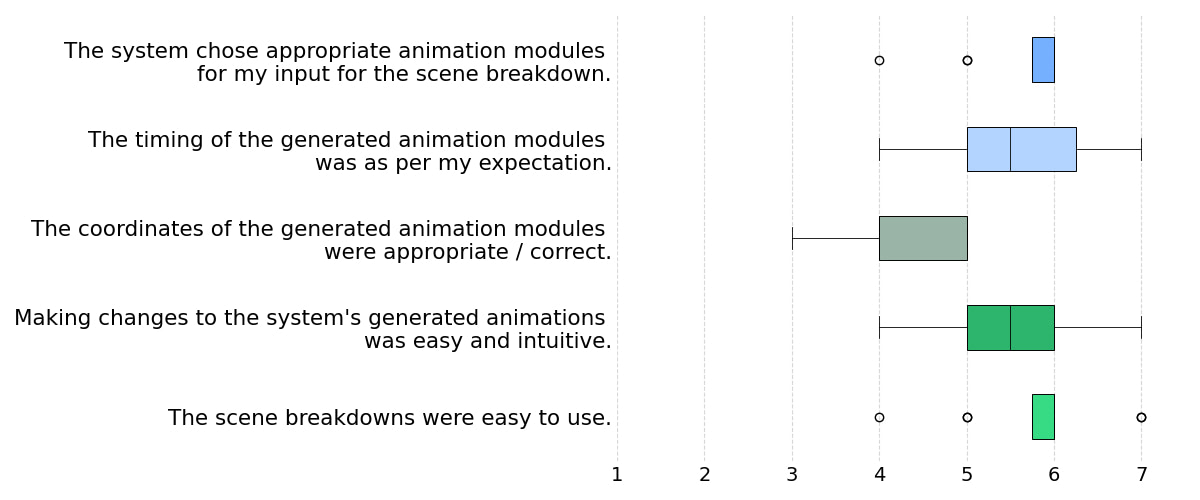}

\caption{Results of the questionnaire for each system feature. The Likert scale ranges from 1 (strongly disagree) to 7 (strongly agree). Outliers (represented by circles) were defined as values that lie beyond 1.5 times the inter-quartile range (IQR) from the quartiles.}
\label{fig:usefulness}
\end{figure}

\subsubsection*{\textbf{Accuracy, Usability, and Usefulness}}
\begin{figure}[h!]
\centering
\includegraphics[width=\linewidth]{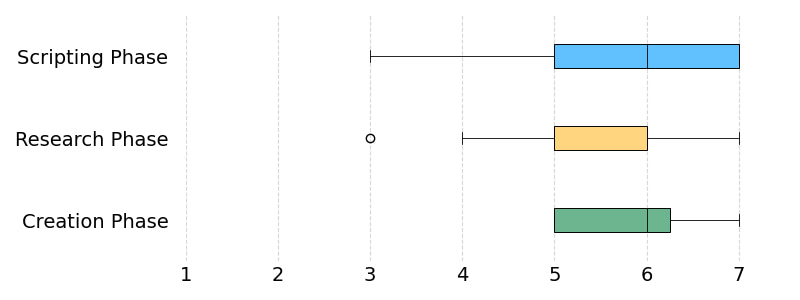}

\caption{Results of the perceived usefulness of each work phase. The Likert scale ranges from 1 (not useful at all) to 7 (extremely useful). Outliers (represented by circles) were defined as values that lie beyond 1.5 times the inter-quartile range (IQR) from the quartiles.}
\label{fig:usefulness_each}
\end{figure}

Overall, the system generated a total of 114 animation modules during our user evaluations. Of those, participants rated 91 of them to be accurate (79.82\%). \textit{Task A} was found to be 89.47\% accurate (34 / 38) and the failure cases were the result of users not being satisfied with the start and end position, which they then prompted the system again to change (After the user changes, the Task A success rate was 100\%). \textit{Task B} was found to be 75\% accurate (57 / 76). Failures were almost always a result of the route or the highlight modules having imprecise coordinates, albeit they were typically within 1km of the correct result. For example, when users asks to zoom into \textit{``a random seven eleven in japan``} the system returned a highlight around 1km away from a seven eleven. Another interesting find was for compound geojson queries like \textit{``the roman empire``}, the system returned the correct regions but some regions were fully highlighted as returned by the present day boundaries, whereas they should be represented with partial borders. In terms of usability, the average System Usability Scale (SUS) score was 83.7 (SD=6.9), indicating high usability. 
As shown in Figure~\ref{fig:usefulness}, system features were rated positively in terms of both accuracy and usability. This indicates that participants found the system features to be reliable and easy to use.
In terms of perceived usefulness, the average scores were 5.75 (SD=1.29) for the scripting phase, 5.33 (SD=1.15) for the research phase, and 5.83 (SD=0.83) for the creation phase, demonstrating high perceived usefulness across all phases (Figure~\ref{fig:usefulness_each}).

\subsubsection*{\textbf{Creativity support by providing scaffolding}}

\begin{figure}
\centering
\includegraphics[width=1\linewidth]{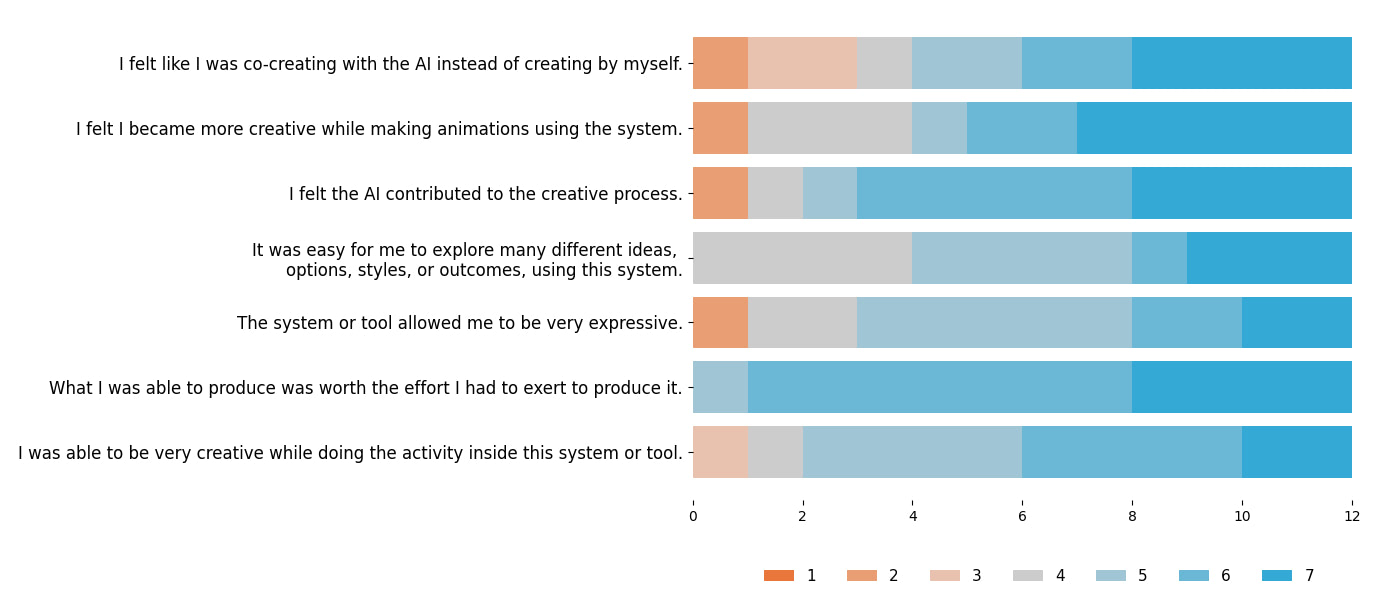}
\caption{Results of the Creativity Support Index (CSI). The Likert scale ranges from 1 (strongly disagree) to 7 (strongly agree). Each bar in the chart represents the number of participants who selected each corresponding response value for the CSI items.}
\label{fig:csi_results}
\end{figure}

Figure~\ref{fig:csi_results} presents the results of the Creativity Support Index (CSI). As shown, participants generally felt that MapStory supported their creativity in addition to enabling them to create map animations, a task which they could not see themselves doing before. This perception was supported by interview data, which suggested that the system lowered initial barriers to entry by providing them with an initial \textit{``guide``} or a starting point. \textit{"In the beginning, it's probably hard for me because there's no starting point, and it's hard to imagine. Everything is like drawing a picture without a draft. But once you have the draft, everything seemed pretty clear, you know, like the immediate next step."} (P9)
Additionally, by decomposing the animation process into manageable steps, the system enabled more efficient prototyping and iteration, which participants found particularly beneficial. 
\textit{"A prototyping tool for different shots would be really interesting, and very quick to do actually."} (P6)
These findings suggest that MapStory not only makes animation creation more accessible by reducing the cognitive load of starting from scratch but also provides structural support for sustained creative exploration and refinement.







\subsubsection*{\textbf{Emerging Use Cases}}
Participants envisioned novel and often informal use scenarios that went beyond our initial expectations, particularly highlighting casual and presentation-oriented applications outside educational or learning video creation. Notably, 8 out of 12 participants proposed using the tool to plan and share travel itineraries in a visually engaging way.
\textit{"If I’m going on a trip or a vacation, I want to share with others, in a fun way, the route I’ll be taking and where we’re going. It sounds a lot more exciting than just giving them an Excel sheet with all the details"} (P10).
\textit{"If you were a travel agent, you could use it to visually present your proposed plan to clients"} 
 (P6).
Also, participants proposed personal documentation scenarios. For instance, one participant mentioned using the tool to visually record their visits to restaurants around the city.
\textit{"I can also use that to document all the restaurants that I've been to around the city"} (P10).
Beyond video creation, the tool was seen as a potential resource for real-time classroom presentations.
\textit{"This would be a great educational tool. Like, if I were a teacher, I could demonstrate it to the whole class"} (P9).
These unexpected use cases highlight the flexibility of the tool and its potential to support a wide range of applications beyond our original scope.

\section{Expert Interviews}
To assess \system{}'s value and feasibility in professional workflows and gain insights into potential use, we conducted expert interviews with five professional map animators (N=5), each with 3–15 years of experience using tools like Adobe After Effects and Davinci Resolve. 

\subsection{Method}
We recruited five professional map video creators (5 male, ages 22-38). Two of these experts had also taken part in our earlier formative study. Each participant had at least two years of experience (2-10 years) producing map animation videos. Three out of the five experts regularly upload videos to YouTube, while the remaining two primarily do client work. Collectively, their Youtube channels have an average of 120k subscribers (5k-300k as of March 2025). We contacted them via Youtube or Instagram, conducted semi-structured interviews of approximately 60 minutes over Zoom, and compensated each with \$50 USD. Experts were given a demo of MapStory and asked to use the system to prototype animations based on a familiar script. We followed up with questions to understand how MapStory could potentially integrate into their workflows.

\subsection{Insights and Findings}
Experts found script-driven animation intuitive and were satisfied with animation controllability, though several noted a desire for even more fine grained control available in professional software. We condense insights from these interviews below.

\subsubsection*{\textbf{\textit{Scene breakdown as the spine of story}}}
Participants found that the scene breakdown not only provides animation guides but also acts as a medium of story exploration. Animators felt that they could stay in \textit{“script mode”}, using short text edits to restructure the whole animation instead of dragging keyframes or rewriting code, while visualizing how the story is presented in real‑time. E1 explained that the scene breakdown provides an initial guided path, which \textit{“is effectively the basics of a story”} and keeps narrative flow intact, and he can directly control and change the narrative: \textit{“The auto‑generated ‘scene breakdown’ feels like \textbf{way‑points in your story that you can tweak in plain language}.”} E5 finds the scene breakdown to be a medium of updating story beats first for editing the visuals. He explains that he traditionally wrote scripts in \textit{“Then/but/therefore”} chunks and then assigns timecodes to make animations in After Effects. But now that he can directly make/write those beats, he can spend his creative time adjusting nuance instead of laying every keyframe. He says, \textit{“[MapStory] generates those things, the clips, the timeline items, and then you can manipulate it later to fine‑tune what you’re doing; that’s the most beneficial.”}

\subsubsection*{\textbf{\textit{Animation by means of conversation}}}
Experts described the chat‑based Researcher as a dialogue that lets them refine map animations and data with text instead of keyframes. E1 explained that when a border is not up to their linking they can \textit{``just ask for information and not just animation,``} saving them from \textit{``manually rearranging keyframes or remaking the scene by hand``}. E5 echoed the sentiment, saying \textit{``AI can take out the grunt work, but then people can tweak the results to a really high degree afterwards``}. E3 quantified the gain: being able to chat‑edit a mock‑up “could save me like 40\% of the time”, and, when modern polygons include unwanted islands, “being able to iteratively change it for my liking is the best part”. Across interviews, conversational animate instructions—whether querying historical borders, trimming overseas territories, or requesting fresh coordinates or information, were praised as a faster, more flexible alternative to the traditional cycle of swapping between Wikipedia, Geo editing tools like QGIS and After Effects.

\subsubsection*{\textbf{\textit{Iterative and editable animation authoring}}}
Creators repeatedly framed MapStory as a way to \textit{rough‑cut first, polish later}, using fast textual edits to explore ideas and postpone heavy key‑framing until they are sure of the direction. E1 calls the system a \textit{quick tool that still gives high quality} output—good enough to visualize and use for final polish—and estimates it can \textit{shave off about two days on a typical five‑day timeline} because the early prototyping phase no longer stalls on full‑resolution renders. The creators still preferred to polish the final video to their liking with advanced editing tools that offer precise control. E5 values the fact that MapStory can \textit{generate the timeline items and then manipulate them}, which he says is \textit{the most beneficial} part of his workflow: MapStory lets him iterate on narrative story beats before committing to detailed animation \textit{``The ability to manipulate is paramount``}. E1 notes that polygon edits are now more approachable and iterative instead of time consuming of GeoJSON changes: \textit{It does these computations by itself … searches for everything and compiles it for you.} Even though the system wasn't perfect, he still appreciated ability to keep prompting or manually make changes \textit{``I really like how you can ask for information and not just animation``}. E2 contrasts this with his current After Effects practice where, \textit{once you make it, you can’t change it… But [with MapStory] I can change it, but it will be time‑consuming}. In MapStory he simply deletes a label and replaces a location highlight and the highlight and camera jumps from India to France, enabling many quick versions: \textit{In 20 or 10 minutes I can do this kind of base animation… then later move it into After Effects.}

Speed also opens space for parallel ideation. E4 keeps several prototypes open at once: \textit{You can do like three videos at the same time … it handles all animation by itself so you don’t have to worry about it.} Because the Researcher answers questions and returns edits, E2 expects to brainstorm more adventurous storylines, noting that the tool itself becomes a live fact‑checking partner. But he also mentions that this would only benefit him during time crunches, and that he would want more precise control (like After Effects) for other projects \textit{``It can’t replace After Effects… your product’s job is mapping, idea tool for a client demo.``}.  Because each animation block is stored as editable prose, creators immediately thought about portability: E3 asked, \textit{Could you save the specific timeline as a file and then import it later?}, treating the breakdown like an edit‑decision list that can hop between tools, while E2 imagined the same promptable timeline living inside Adobe’s flagship: \textit{If we integrate this with After Effects, then I would love [to use it] … it gives me reassessing thing.} Text‑driven prototyping lowers the cost of experimentation, lets animators iterate early and often, and reserves advanced tools for the final layer of visual details.

\subsubsection*{\textbf{\textit{Limitations and caution}}}
Several experts flagged the risk that an LLM‑powered “Researcher” might invent or distort facts. E4 warned that prompts can return \textit{``wrong information or … hallucinations … I will double‑check``} before publishing anything. Yet most participants stressed that fact‑checking is not new creators have always had to validate sources. E5 summed it up bluntly: \textit{``You can’t trust Wikipedia … you’ve gotta do a lot of research to make sure things are historically accurate, a lot of stuff out there is just wrong.``} . Likewise, E3 treats AI outputs like any other draft, noting that he needs a “general idea” so he can spot errors and avoid \textit{``accidentally including the AI mistake``} in his maps. Participants did appreciate the controllability our system provides. Moreover, the experts explained that they would not find themselves using MapStory as their final production tool but more for initial exploration, drafting, or pre-production for client demo or storyboarding. They prefer having complete creative control over their animations with advanced software which support keyframing like After Effects or Davinci Resolve.

%% file: 8-future-work.tex
\section{Future Work}

\subsubsection*{\textbf{\textit{Modular extensibility of AI‑assisted, text‑driven animation prototyping}}}  
Our formative explorations have yielded a corpus of map‑centric animation blocks; however, our modular system can be extended with new, custom blocks tailored to animators’ niche tastes. Participants envision creating their own stylistic modules—such as filters, transitions, or bespoke map‑animation blocks, while some imagine importing a hand‑designed basemap and letting the engine treat it as another selectable style, \textit{``Imagine you can import … an image, and then it would do the same thing``} (E3). Because each block is merely metadata plus a renderer hook, contributors can publish plug‑ins (e.g., “smear‑frame transition,” “parchment basemap”) to a shared registry. Experts also envision a community forming around trading map‑specific animation blocks, basemaps, transitions, and filters, similar to in‑game items.

\subsubsection*{\textbf{Geocoding Ability of LLMs}}

Our system is constrained by how accurately LLMs can map textual descriptions onto geographic coordinates and generate valid GeoJSON (albeit geocoding accuracy improvement was never our goal). Although our functions give animators interactive, iterative control, reliable geocoding remains challenging. For example, when visualizing the Roman Empire we can highlight its key regions, but we must rely on present‑day boundaries; portions of the historical empire correspond to only parts of modern countries—a level of detail our system cannot yet represent (users can, however, draw a rough outline manually). Despite this limitation, the text‑driven animation workflow itself remains unchanged. We also acknowledge that our evaluation was restricted by API rate limits, which limited the number of prompts we could test. Future work should therefore conduct a more extensive assessment focused on geocoding to explore the limits and capabilities of LLMs for any map‑based visualization, not just animations. Our agent design is also limited by a design features two agents. we chose this design based on our formative evaluations and time constraints of animation authoring. However, future architectures could explore agents interacting over multiple rounds to keep refining the geocoded result of the previous agent until a verification agent is satisfied.

\subsubsection*{\textbf{Beyond Maps: Broader implications of text or conversation driven animation workflows}}
The interviews suggest that MapStory’s text‑driven “scene breakdown + Researcher” model could serve as a control layer for broader animations, not just maps. Once the timeline itself is conversational, the Researcher can query far more than geographic data; E5 already envisions layering non‑map treatments—\textit{``You can also add transitions or filters to the maps to give it more [board‑game] kind of a thing.``} —and the same mechanism could contextually search a studio’s library of transitions, LUTs, or particle systems, insert the chosen effect blocks into the timeline (additionally add transitions or filters based on user's intention), and expose their parameters for manual tweaking at every stage. This can create a reusable, domain‑agnostic interaction that can accelerate storyboarding and live client iteration for any genre of animation.

%% file: 9-conclusion.tex
\section{Conclusion}
This paper introduced \system{}, a text-driven map animation prototyping tool that enables users to create map-based animations through natural language editable at every stage. We contributed to the design space and LLM architecture of our system. Formative evaluations detailed workflow of map animation creation with a design space of map-animation blocks. \system{} introduces modular animation blocks, integrated geospatial querying researcher, and tight coupling between script and animation through a step-by-step scene breakdown. These features were implemented through novel architecture that leverages two LLM agents, one for scene breakdown and another one for research agent. Through technical evaluations, usability studies, and expert interviews, we found that \system{} supports fast iteration, encourages creative exploration, and aligns well with existing professional workflows. On the other hand, our findings also revealed several limitations, including occasional inaccuracies from LLM-generated outputs and needing verification and finer grained control. Future work should address these limitations to enable more accessible, expressive, and verifiable map animations.

%% file: acknowledgements.tex
\begin{acks}
This research was partially funded by the NSERC Discovery Grant RGPIN-2021-02857, JST PRESTO Grant Number JPMJPR23I5, and Adobe Collaborative Research Gift. We also thank all of the participants for our user study.
\end{acks}

%% file: videosource.tex
\section*{List of Image Sources}
In Figure 2 and 4 we have used still frames from videos published by the following creators on YouTube.
\onecolumn
\begin{longtable}{p{6cm} p{10cm}}
  \caption{YouTube Video Frame Sources}\label{tab:yt-sources}\\
  \toprule
  \textbf{Copyright Name} & \textbf{Channel Link} \\
  \midrule
  \endfirsthead

  \multicolumn{2}{l}{\small\sl continued from previous page}\\
  \toprule
  \textbf{Copyright Name} & \textbf{Channel Link} \\
  \midrule
  \endhead

  \midrule
  \multicolumn{2}{r}{\small\sl continued on next page}\\
  \endfoot

  \bottomrule
  \endlastfoot
  © Copyright Ahmed Sabry      & \url{https://www.youtube.com/channel/UCpLQ1u9_KSTVAO3PXkJho4w} \\
  © Copyright Atlas Pro        & \url{https://www.youtube.com/channel/UCz1oFxMrgrQ82-276UCOU9w} \\
  © Copyright BBC News         & \url{https://www.youtube.com/channel/UC16niRr50-MSBwiO3YDb3RA} \\
  © Copyright Binkov's Battlegrounds & \url{https://www.youtube.com/channel/UCPdk3JuQGxOCMlZLLt4drhw} \\
  © Copyright Business Insider & \url{https://www.youtube.com/channel/UCcyq283he07B7_KUX07mmtA} \\
  © Copyright CGP Grey         & \url{https://www.youtube.com/channel/UC2C_jShtL725hvbm1arSV9w} \\
  © Copyright CaspianReport    & \url{https://www.youtube.com/channel/UCwnKziETDbHJtx78nIkfYug} \\
  © Copyright Cycle V          & \url{https://www.youtube.com/channel/UCaCuAz2fXGVWMEt4jWUtTCA} \\
  © Copyright EmperorTigerstar & \url{https://www.youtube.com/channel/UCUXqYwTCR6R3Wr-FkLTD4AQ} \\
  © Copyright Epic History     & \url{https://www.youtube.com/channel/UCvPXiKxH-eH9xq-80vpgmKQ} \\
  © Copyright FactSpark        & \url{https://www.youtube.com/channel/UC5EcQrgvXsgGLR4ukxBTM8w} \\
  © Copyright Geo All Day      & \url{https://www.youtube.com/channel/UCdWuhQg9XRYXTiON98aC03w} \\
  © Copyright Geo History      & \url{https://www.youtube.com/channel/UC2Cl2g2xFTZoAEldxYVzQFg} \\
  © Copyright GeoGlobeTales    & \url{https://www.youtube.com/channel/UCPKVNqfbunZCkbvjBgf6CdA} \\
  © Copyright Geography By Geoff & \url{https://www.youtube.com/channel/UC7-Z34pbJ0ZAOJRmUCEz0Cg} \\
  © Copyright Globe Addict David & \url{https://www.youtube.com/channel/UCXJniNhLPJSeYXAfUzexHSA} \\
  © Copyright Histoire Géo     & \url{https://www.youtube.com/channel/UCYP6V7c4M026AXMr4ng6aQA} \\
  © Copyright History Mapped Out & \url{https://www.youtube.com/channel/UCtzvIHQyRDL2mtetu6ZWsvw} \\
  © Copyright History Matters  & \url{https://www.youtube.com/channel/UC22BdTgxefuvUivrjesETjg} \\
  © Copyright History On The Map & \url{https://www.youtube.com/channel/UCYNb4d5zyaJ6KIN9knzoevg} \\
  © Copyright History Scope    & \url{https://www.youtube.com/channel/UCYb6v1AlX6prKl83DswM5iw} \\
  © Copyright History on Maps  & \url{https://www.youtube.com/channel/UC7JrMLRDJLUh5BPBYykMcrQ} \\
  © Copyright HistoryMarche     & \url{https://www.youtube.com/channel/UC8MX9ECowgDMTOnFTE8EUJw} \\
  © Copyright Jabzy            & \url{https://www.youtube.com/channel/UCoUkea_dZioNSJbi1vWDZkA} \\
  © Copyright Jay Foreman      & \url{https://www.youtube.com/channel/UCbbQalJ4OaC0oQ0AqRaOJ9g} \\
  © Copyright Johnny Harris    & \url{https://www.youtube.com/channel/UCmGSJVG3mCRXVOP4yZrU1Dw} \\
  © Copyright Kings and Generals & \url{https://www.youtube.com/channel/UCMmaBzfCCwZ2KqaBJjkj0fw} \\
  © Copyright Knowledgia       & \url{https://www.youtube.com/channel/UCuCuEKq1xuRA0dFQj1qg9-Q} \\
  © Copyright LEMMiNO          & \url{https://www.youtube.com/channel/UCRcgy6GzDeccI7dkbbBna3Q} \\
  © Copyright Look Back History & \url{https://www.youtube.com/channel/UCnPYviDHmuqLeKG6kfLTV8w} \\
  © Copyright Map Pack         & \url{https://www.youtube.com/channel/UCb0JxFSwHTzH0AEsAzQHV4w} \\
  © Copyright Megaprojects     & \url{https://www.youtube.com/channel/UC0woBco6Dgcxt0h8SwyyOmw} \\
  © Copyright Money \& Macro   & \url{https://www.youtube.com/channel/UCCKpicnIwBP3VPxBAZWDeNA} \\
  © Copyright Odd Compass      & \url{https://www.youtube.com/channel/UCpZYcHFVksxNCwrUKgJMDyw} \\
  © Copyright Ollie Bye        & \url{https://www.youtube.com/channel/UC6gNjP1W4FXWExT5QpYkmhQ} \\
  © Copyright OnlyIAS Extended & \url{https://www.youtube.com/channel/UCAidhU356a0ej2MtFEylvBA} \\
  © Copyright OverSimplified   & \url{https://www.youtube.com/channel/UCNIuvl7V8zACPpTmmNIqP2A} \\
  © Copyright Overly Sarcastic Productions & \url{https://www.youtube.com/channel/UCodbH5mUeF-m_BsNueRDjcw} \\
  © Copyright PBS Eons         & \url{https://www.youtube.com/channel/UCzR-rom72PHN9Zg7RML9EbA} \\
  © Copyright RealLifeLore     & \url{https://www.youtube.com/channel/UCP5tjEmvPItGyLhmjdwP7Ww} \\
  © Copyright Server           & \url{https://www.youtube.com/channel/UCY_T0aB9ch-oYv7tBSuTC_w} \\
  © Copyright TED-Ed           & \url{https://www.youtube.com/channel/UCsooa4yRKGN_zEE8iknghZA} \\
  © Copyright TLDR News Global & \url{https://www.youtube.com/channel/UC-uhvujip5deVcEtLxnW8qg} \\
  © Copyright The Generalist Papers & \url{https://www.youtube.com/channel/UCN9UPjA8I-uwvAy0-N9maOA} \\
  © Copyright The Wall Street Journal & \url{https://www.youtube.com/channel/UCK7tptUDHh-RYDsdxO1-5QQ} \\
  © Copyright Theology Academy – Christianity & \url{https://www.youtube.com/channel/UCq7kcinj-b9d57THam2QOBQ} \\
  © Copyright True Story Docs   & \url{https://www.youtube.com/channel/UCALxmR5lDHHjN25Fb0ODZdw} \\
  © Copyright Uncovering       & \url{https://www.youtube.com/channel/UCrsm1dqSALPG5BCCiI0Ebug} \\
  © Copyright Virtual High School & \url{https://www.youtube.com/channel/UCveQN5D9y28Xou9jMBQkbGQ} \\
  © Copyright Vox              & \url{https://www.youtube.com/channel/UCLXo7UDZvByw2ixzpQCufnA} \\
  © Copyright WeatherWatchTV   & \url{https://www.youtube.com/channel/UC6J_1lbR727yUrcm9m140Ug} \\
  © Copyright Wendover Productions & \url{https://www.youtube.com/channel/UC9RM-iSvTu1uPJb8X5yp3EQ} \\
  © Copyright WonderWhy        & \url{https://www.youtube.com/channel/UCcEPmwpXKrKzZahqjwpIAsQ} \\
  © Copyright pigeoninanutshell & \url{https://www.youtube.com/channel/UCQQwkFEwF4TfyOr26Yu6-vA} \\

\end{longtable}